# Motoric signatures to detect original authorship versus cognitive effort in instructed motor responses of an emerging speaker with ASD

(*Dedicated to Anne M. Donnellan, 1943-2024*)

*Nothing in life is to be feared, it is only to be understood. Now is the time to understand more, so that we may fear less.* — Marie Curie

**Elizabeth B Torres**[1,2,3]

[1]Sensory Motor Integration Laboratory, Psychology Department, Rutgers the State University of New Jersey, New Brunswick, NJ, USA

[2]Rutgers University Center for Cognitive Science, Rutgers the State University of New Jersey, New Brunswick, NJ, USA

[3]Center for Biomedicine Imaging and Modelling, Computer Science Department, Rutgers the State University of New Jersey, New Brunswick, NJ, USA

*** Correspondence:**
Elizabeth B Torres
ebtorres@rutgers.edu

**Abstract**

There is at present no objective, agnostic method to track levels of message authorship and agency in non-speakers who rely on assisted typing to communicate. Part of the problem is that nuances in spontaneous gestural movements of the typer transpire largely beneath awareness of both the typer and the communication-regulation-partner. Such spontaneous movements co-occur with the intentional ones producing the typed messages but are currently not considered part of the communication system. We introduce an assay that probes various levels of cognitive loads in an ecologically valid environment. Then, in collaboration with an emerging speaker-typer and his communication-regulation-partner, we empirically estimate motoric stochastic signatures during both independent and assisted typing. Under varying levels of cognitive loads, the assay and biometrics distinguish motoric patterns in spontaneous typing of phrases in response to open-ended questions from those in instructed typing of the same phrase. The non-dominant bodily motions, co-occurring with the dominant (typing) side reveal not only levels of cognitive mental effort and agency of the typer but also levels of creative thoughts through motoric preambles present only in open-ended mode. This level of semantic authorship is detectable in both the geodesic paths along the manifold representing the Gamma stochastic parameters of the motions' fluctuations and those of the physical 3D spatial trajectories. The head and facial micro-motions capture these non-dominant-body nuances, offering a personalized motoric metric of cognitive effort, agency, authorship and a comprehensive suite of assays to open new research avenues for assisted typing with haptic support.

## 1 Introduction

Communication is paramount to human agency. The ability to immediately transmit, at will, our feelings, emotions, and mental or physical states to others can be a life saving grace in difficult situations. In the spectrum of autism (ASD) communication is only recently being addressed and no proper body of research exists to guide the process of language acquisition when neurodevelopmental paths leading to autonomous motor control are delayed or stalled, as it seems to be the case in the type of ASD that requires high levels of support and cannot speak [1].

Conservative estimates by the US Centers for Disease Control (CDC) suggest that approximately 30% of autistic people are minimally verbal or non-speaking (ns-ASD) [2, 3]. There is however no reliable count or statistic informing us how many people can type or point to communicate. In advanced cases, the autistic person communicating with others can fluidly navigate from a keyboard to a tablet, to a letterboard (**Supplementary Figures 1-3**). Most importantly, advanced users of these techniques can utter the words being typed, as the communication regulating partner (CRP) becomes merely a presence to help regulate the person's body as needed. One way to regulate the typer is through haptic feedback in the form of light touch. Those few ns-ASD who learned to type independently and even began to vocalize the phrases they type, followed a learning trajectory that brought them to that level of independence. Yet we do not understand the mechanisms that led to this extraordinary feat. We need metrics that help us track that process in ns-ASD, particularly when we know that neurotypical language acquisition follows a very different, inverted neuromotor path first learning how to talk and then later, by school age, mastering writing/spelling to communicate. Research addressing the ns-ASD learning process is missing.

As learning to communicate is a process that takes years, it evolves through different stages of autonomy. These may initially rely excessively on the CRP but as learning milestones are achieved, the CRP gradually transfers agency to the typer. The average ns-ASD engaging with these adopted world-wide communication techniques needs the regulating partner to keep the conversation flowing. Along those lines, recent advances in biometrics from digital data reveal not only the possibility of tracking the progression of agency while acquiring independence [4] but more importantly, they reveal the regulation levels badly needed in the process of learning to type independently (See **Supplementary Figure 4**).

Research involving motor control metrics derived from wearables' kinematic data demonstrated that touch may reduce cognitive load during assisted typing [5]. They have also demonstrated, using corpora of 20 years of communication involving multiple CRPs, that ns-ASD using these techniques make their own stylistic contributions to text written with such physical facilitation [6]. And eye-tracking studies in the US suggest that ns-ASD participants predictively look first at the letter that they are about to point during spelling exercises [7]. Studies from Israel suggest that saccades in ns-ASD may demonstrate receptive language better than pointing (which is more arduous) [8], despite excessive random noise in their eye motions [9]. This result concerning the eyes' motions is congruent with excessive random noise found in their involuntary head motions at rest [10], their erratic heart rate variability in sympathetic overdrive, even at rest [11-14], their dysregulated gait [15] and their voluntary movements aimed at visual targets [16, 17] or when trying to orient to and/or differentiate auditory targets [18-20]. **Supplementary Figure 4** offers compelling evidence for the need to help regulate the autistic system and move it towards physical comfort.

### 1.1 Regulating Motor Control by Offering a Frame of Reference

Physical independence, however, is different from authorship of communicating messages. The latter can happen even when the person is being physically supported to provide anchors across the body and help create interchangeable *frames of reference*. Frames of reference are necessary in sensory-to-motor coordinate transformations because different extrinsic and intrinsic sensory sources have disparate latencies in signals' transduction and transmission across different brain-body systems.

When a visual target (such as a letter on a keyboard or letter board) is presented to the brain, or an auditory cue is used to prompt a goal-directed action, the head must properly orient and coordinate with the trunk and the hand to deliver a fluid pointing movement to the targeted letter. Several sensory signals' transductions, transmissions and transformations delays occur in the process of delivering the motor commands, before the hand is launched towards the target in a controlled and well-coordinated manner. The brain must transform signals, from retinotopic to head centered coordinate frames [21]. This flow of sensory information originates in an *external sensory source* (light) and is first mediated by photoreceptors in the eye's retina. To generate goal-directed action to a target external to the body, latencies linked to external sources must align with those linked to *internal sensory sources*, anchored in the body. As the human body has many degrees of freedom (DOF), the unfolding motions afford us great flexibility to accomplish the same final goal in an infinite number of manners – a problem beautifully stated by Nikolai Bernstein in the first half of the last century [22, 23]. This redundancy however, also creates complex computational problems that require time to solve and continuously update during planning and execution of such purposeful actions [24].

To solve these problems, disparate sensory sources must be integrated and transformed into motor coordinates representing the ongoing movements, which are also the returning stream (kinesthetic reafferent feedback [25]) that the brain uses to update and predict impending actions' consequences [24]. Considering their variable latencies is key to timely aligning our actions to our thoughts and to forecasting the impending consequences of both intended and unintended movements.

Kinesthetic reafferent signals mediated by mechanoreceptors coexist with other afferent streams mediated by nociceptive and thermal receptors. These intrinsic signals, coded relative to internal sensory streams, are facilitated in part by the brainstem, a conduit to both subcortical and cortical structures involved in the control, coordination and sequencing of autonomic, reflexive, spontaneous, involuntary, automatic and voluntary motions [26]. Because of its central role, since conception, in central patterns generation and protective reflexes, we posit that the brainstem is key to the chronometric alignment required between external (world-anchored) and internal (body-anchored) sensory sources [27].

Temporal alignment of disparate sensory-motor streams is thus necessary to produce gross motor movements but also to coordinate fine motor control during vocalization via shared structures in the vocal apparatus. There the power source (respiration), the sound source (phonation), and the filter and modifier system (resonance and articulation) must negotiate such temporal alignment codes from birth, to master autonomous control of breathing, swallowing and vocalization [28] -without choking or aspiring food into the lungs while feeding. During neurotypical development, these temporal alignments occur seamlessly so they are obviated in our research programs. However, the moment we interact with autism, we can see prolonged delays in sensory processing and sensory motor integration, since birth [16].

The auditory brainstem response (ABR) test routinely administered at birth in the US, has revealed such prolonged delays in the auditory domain [29-33]. Tracking public records in the states of New York and Florida, we were able to see that neonates with such prolonged ABR latencies and narrowed bandwidth of response ranges at birth, later received the autism diagnosis by 3-4 years of age and they were still present in autistic children between 1-6 years of age [29]. This temporal misalignment of the ABR necessarily impacts the motor code, [26, 34, 35]. The brainstem role in CPGs producing spontaneous movements is key to the maintenance of healthy motor variability necessary throughout life. Healthy motor variability is required to build steady motor percepts and to enable cognitive control of our biomechanical motions [16, 17, 36, 37].

The quality of spontaneous movements in early infancy denotes a healthy neuromotor control trajectory. As spontaneous motions across the body transition into antigravity intentional movements, we can quantify, very early, the proper evolution of voluntary neuromotor control [38, 39]. We can also track the spontaneous motions that are incidental to goal-directed movements [36, 40], as they contribute to our understanding of unintended consequences and help us learn to build awareness through movement [41-45]. Together, through complementary roles, spontaneous and volitional motions can help us track the development of bodily motor percepts, update body schemas and, in the words of Esther Thelen, help us ground and situate the nervous systems for embodied cognition [46-48].

Embodied cognition can be tracked through spontaneous gestural movements embedded in purposeful actions, detectable through biosensors. These include pointing to communicate [16, 49] or performing sports and dancing activities [36]. Their patterns of variability are distinct from those of deliberate segments in the continuous trajectories describing our actions [36, 37, 40]. The spontaneous segments of purposeful movements' excursions described by our end effectors are subject to different geometric constraints than the deliberate segments [37]. Their temporal dynamics reflect the forces that produce our motions [36, 37], unlike the deliberate segments where the geometric features of the intended movement paths tend to remain impervious to changes in the forces [36, 40, 50-52]. Once initiated, geometric features of the intended path to point to a letter on a keyboard remain steady despite changes in the patterns of acceleration and speed. One can therefore automatically identify and isolate key features of our mental intent and key elements of our physical volition in natural motions containing both deliberate and spontaneous segments, [37, 53-57]. Because spontaneous variations in our movements also increase with cognitive effort, they can signal levels of cognitive load [17, 37]. Furthermore, since spontaneous movements flow in open-ended conversations, where they transpire largely beneath the person's awareness, we can use them to understand creative processes in the production of semantic phrases to communicate our thoughts. Spontaneous movements are spared in autism [16]. They thus offer a malleable, plastic signal to help the system develop the sense of action ownership and ultimately enhance the sense of agency in the person [17]. As such, we can use spontaneous motions to help us track levels of agency and ownership of the person's inner thoughts as they unfold in assisted communication.

In this paper, we define a suite of assays and personalized, standardized biometrics that leverage the spontaneous movements of the non-dominant side of the body to distinguish authorship in open-ended vs. instructed communication settings. As a person types or gestures to communicate with the dominant side of the body, the non-dominant side simultaneously moves largely beneath awareness of the typer and the CRP. Consequently, motions of the non-dominant side co-occurring with those of the dominant side during communication, cannot be deliberately manipulated by an external agent without being noticed. We find that the head motions are closer to the motions of the non-dominant side of the body, both in physical and stochastic spaces, thus offering the potential to serve as a proxy broadcasting levels of agency, cognitive effort and semantic authorship. Here we provide agnostic biometrics to objectively track the course of communication in a person, both when the person types independently and when the communication is assisted by haptic feedback. We offer a new avenue to research communication in the face of somatic-sensory-motor control issues demanding bodily support.

## 2 Methods

### 2.1 Participant

This study took place at the Sensory Motor Integration Laboratory of Rutgers University under IRB protocol Pro2020000154 ("*Digital Characterizations of Nervous Systems*") in compliance with the Helsinki Act. We attained consent from the adult participants to be videotaped and photographed. This is a case study involving Matthew Hayes (MH) an emergent speaker adult with a diagnosis of ASD who willingly participated and communicated with our team using various means. These included iPhone, pocket-size cardboards with letters, larger letterboards and electronic keyboards. The participant could type independently (without the keyboard being sustained by another person). As he typed, he repeated the words out loud. At the end of typing the full phrase, he said it out loud as well. In some instances of the suite of assays that we designed, the CRP provided haptic feedback by light touch of the right knee to dampen involuntary motions of the lower body. To sooth himself, the participant would spin a hanger with the typing hand when he was not typing. As he set the hanger aside to type during the experiments, he produced other wondering motions with the typing hand while thinking about the answers to open ended questions. These wondering motions were minimal during instructed typing, but returned with cognitive effort, albeit in a very different manner than during the expansive open-ended motoric preamble. We took breaks between conditions to minimize and avoid fatigue. The participant indicated when he wanted to break to drink water, snack or simply rest from the experiment.

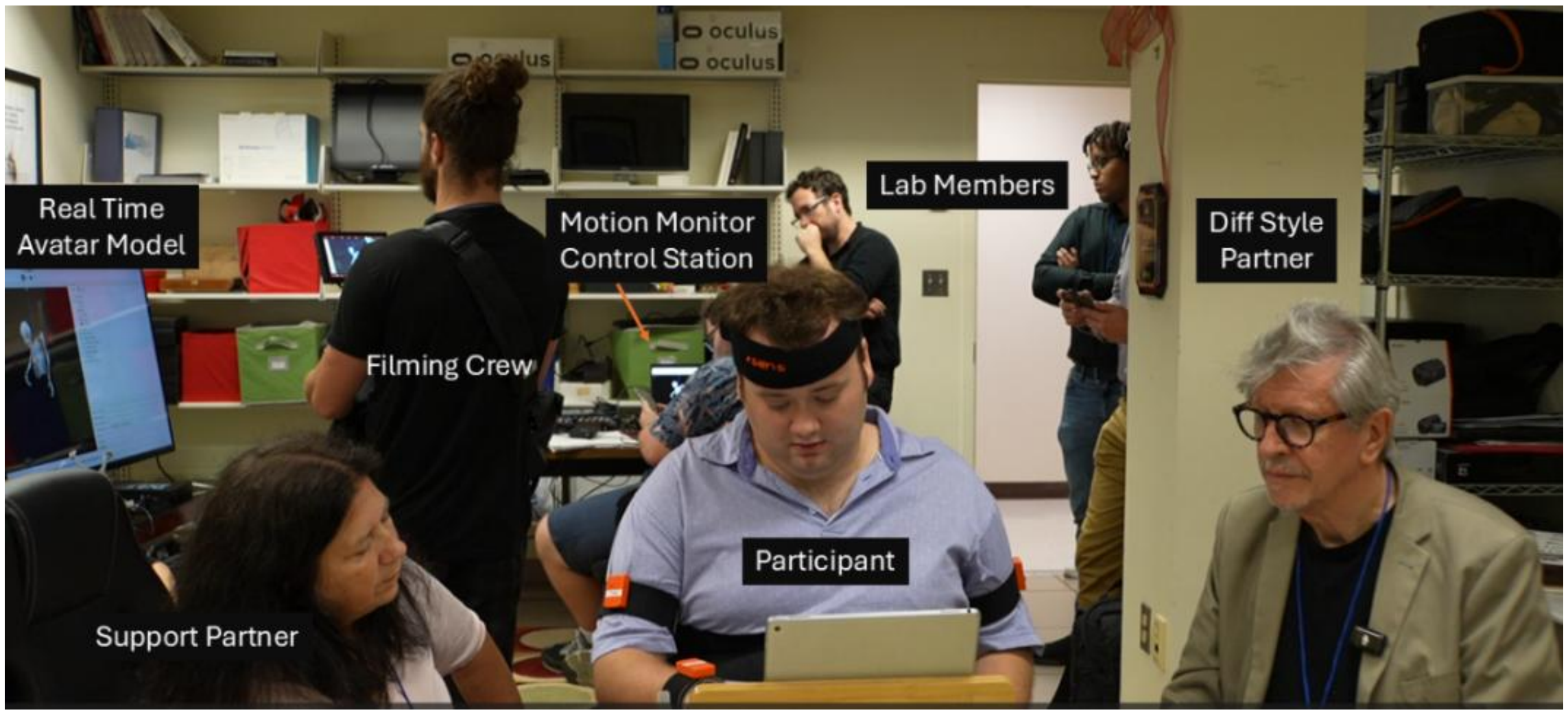

**Figure 1. Experimental apparatus and participants. Matthew Hayes (MH) and his mom, Rene Hayes sat next to each other facing a keyboard and a screen on a stand, where the output of the typed words could be seen. Dr. Daniel Orlievsky observed the performance and later switched to be the communication partner, using a different keyboard on a laptop. The participant had never met him.**

## 2.2 Apparatus

The participant MH sat comfortably by a table, flanked by the CRP and another clinician that he did not know. Across MH sat the first author (EBT) conducting an interview, while two cameras recorded video of the session from different angles. A professional videographer also videotaped across the room. Undergraduate and graduate students assisted with the logistics of the study. Two other clinicians witnessed the interviews. The layout of the experiment is depicted in **Figure 1**.

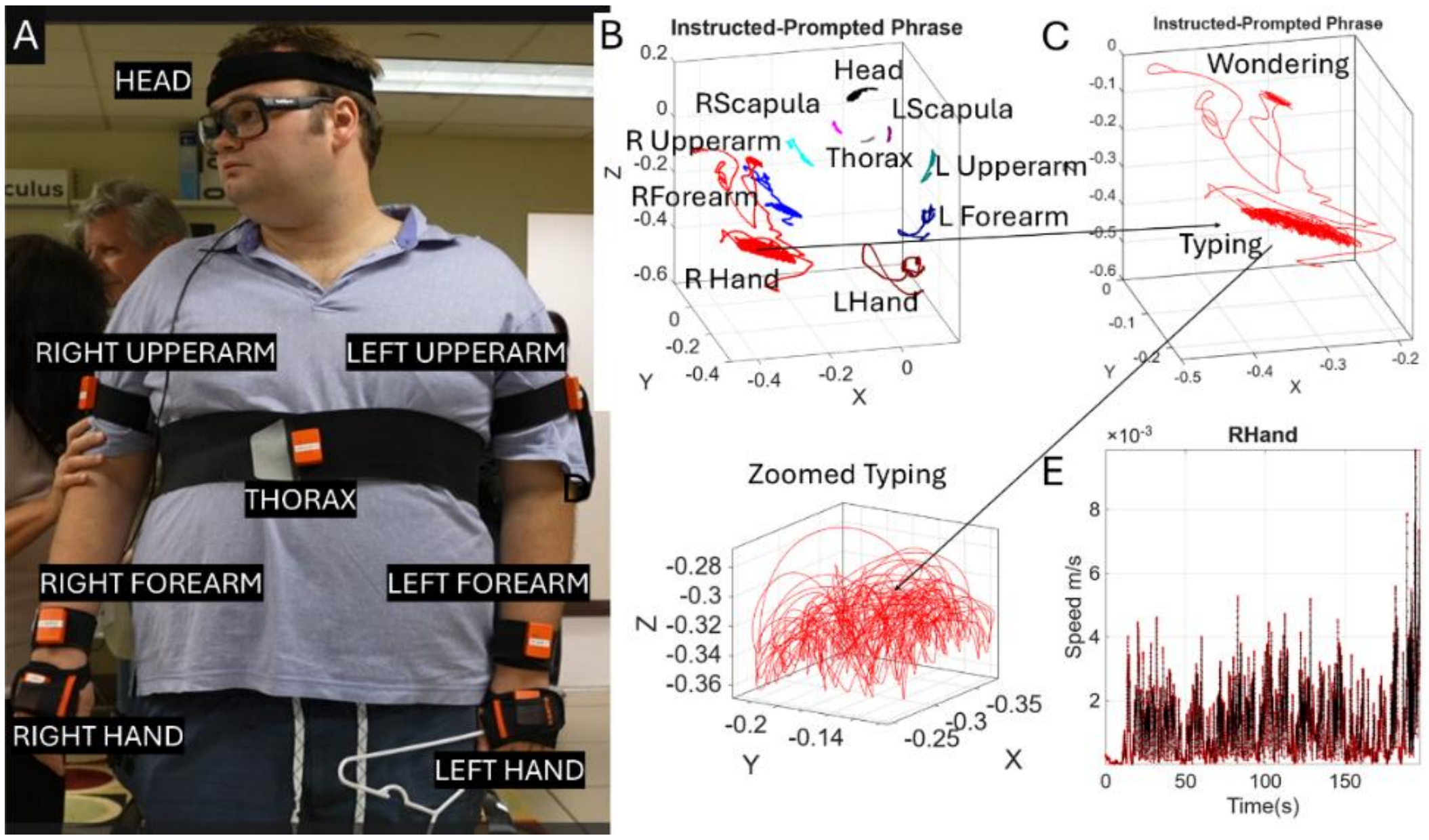

**Figure 2. Sensors and sample trajectories. (A) Eight wearable sensors were used to collect data at 100Hz, located in the shown configuration. (B) Sample 3D positional trajectories from all sensors are plot indicating their location in physical space. (C) Sample hand trajectory reflecting the typing movements and some wondering motions. (D) Zoomed-in hand trajectories from typing motions. (E) Sample temporal speed profile derived from the hand's continuous 3D positional trajectory (smoothed using appropriate in-house developed routines using spline interpolation methods and derivative functions from MATLAB).**

We collected position and orientation data using the XSense system (sampling resolution 100Hz) integrated within the Motion Monitor system (Sports Inn, Chicago). Heart Rate Variability was collected as well with a Noraxon Ultium EMG sensor (sampling resolution 4kHz) synchronously with the kinematics data (down sampled at 100Hz) and within the Motion Monitor interface. We report HRV and face data in a separate paper. A three-dimensional avatar rendering the participant's upper body skeleton displayed the arm joint's trajectories and the head and hands trajectories (end effectors of the upper body) in real time (**Supplementary Figure 5**). Cameras (Sony and Cannon) recorded continuous face and upper body videos for separate analysis using pose estimation. Furthermore, pre- and post-experiment facial scans were taken using the Latently app (Neuroinversa LLC, NJ, USA) running on an iOS tablet. Additional XSense inertial measurement units (IMU) sensors (sampling at 100Hz) were placed on the wrists of the CRP, to further digitize and test other dyadic interactions. **Figure 2** shows the location of the sensors on the upper body and head.

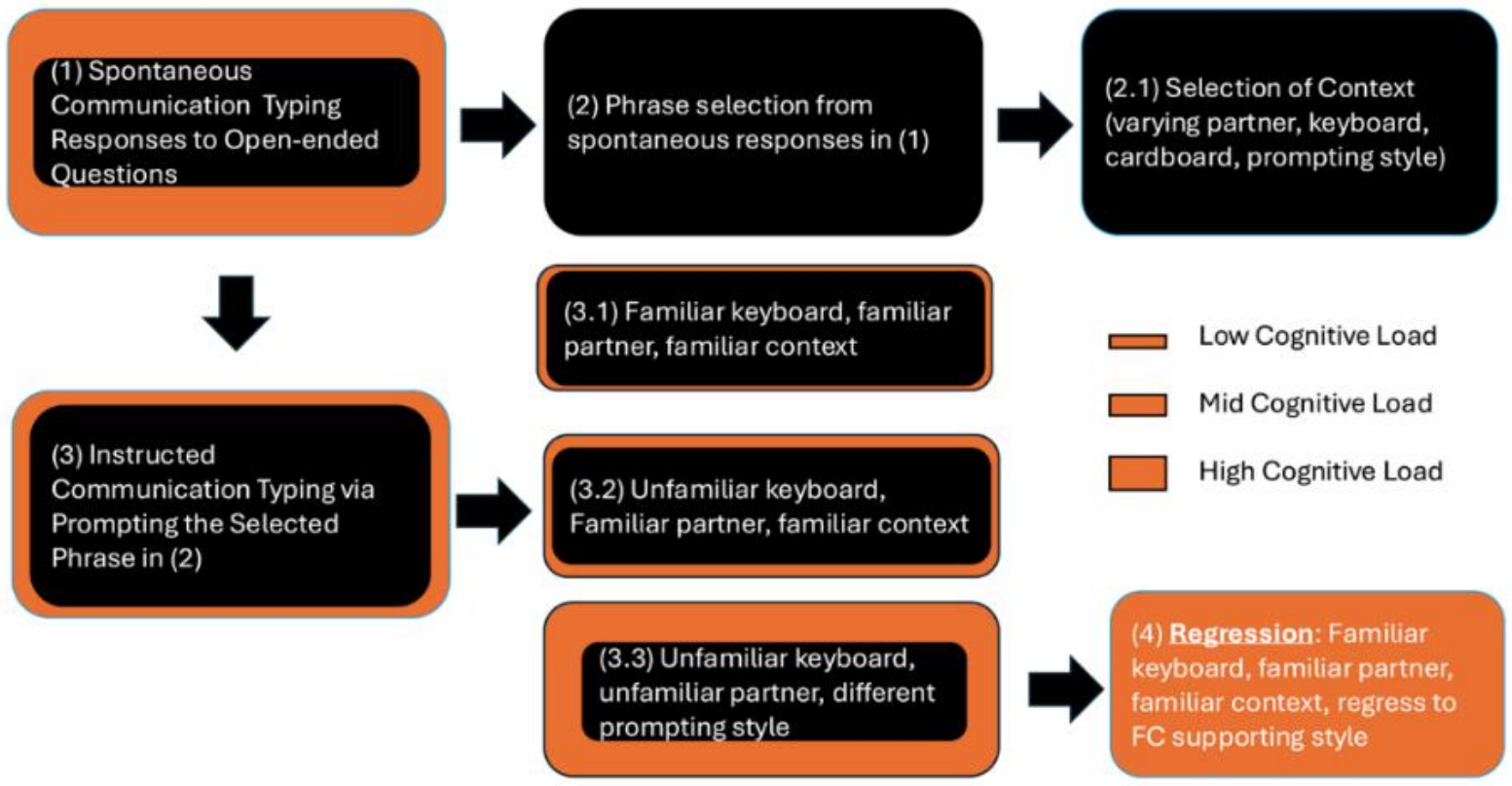

**Figure 3. Assay to automatically extract motoric signatures of authorship in communication through typing. (1) Spontaneous communication is evoked through open-ended questions. (2) A phrase from the answers to open-ended questions is selected to be prompted during instructed typing. (2.1) Variations in the environment are selected to increase the cognitive, memory and emotional load in an ecologically valid manner, like in natural social situations, during the instructed condition. (3) Instructed phrase is typed via prompting the phrase selected in (2). (3.1) Low cognitive, memory and emotional loads during instructed phrase using a familiar keyboard, familiar CRP and familiar context. (3.2) Mid cognitive load using a new (unfamiliar) keyboard but retaining the CRP and similar context. (3.3) High cognitive, memory and emotional loads are introduced by switching to a new, unfamiliar keyboard with different layouts, a different communication partner and different prompting style. (4) Other conditions include covering the keyboard with a white sheet of paper to block view of the letters; regressing to an old-style Facilitated Communication from earlier stages of the learning process, during infancy. Biorhythms from both dominant and non-dominant body locations and end-effectors are co-registered.**

## 2.3 Assay to Automatically Extract Motoric Signatures of Authorship

We examined the biomechanical signatures across the upper body to interrogate fundamental differences between owning the motoric output of the body while elaborating and spontaneously communicating a thought by typing a phrase (and saying it) vs. typing the same phrase while prompted by another person. We use kinematic and stochastic analyses. Results from HRV and facial data will be reported elsewhere. Importantly, we specifically examine the non-dominant side of the body because as the dominant hand types in open ended or instructed mode, or as it types assisted by the CRP, the co-occurring motions of the non-dominant side of the body transpire out of conscious reach from the CRP and from the typer himself. The paradigm is designed to evoke such motoric differences across the following conditions, where order matters (see **Figure 3**).

## 2.4 Warming up motions from Adaptation to a New Context and Environment

Prior to the experiment, we recorded warming up motions. These are spontaneous movements while acclimating to the setup, the sensors, the physical layout of the lab and the social-emotional moods of the group, in general. These movement trajectories registered from the upper body in the physical 3D space along with their stochastic trajectories on the $\Gamma$ manifold were obtained but they are not part of the experiment per se. They are part of the natural adaptation process of any social situation.

## 2.5 The Flow of the Experiment

First, one needs to probe communication performance using open-ended questions that evoke spontaneous thoughts. Throughout this process, we record the phrases typed (and spoken in cases where speech is emerging or steady). In cases where speech is absent, we can record vocalizations, echolalic phrases, self-talk or any spontaneous utterances that the person makes. Second, one needs to select a phrase (vocalization, echolalic or self-utterances) from the spontaneous responses to the open-ended questions or from any spontaneous self-utterances. Finally, probe the selected phrase / self-generated utterances under instruction by prompting it, to type it / spell it under different cognitive, memory and emotional loads. Notice here that launching the arm-hand linkage to type is a gross motor act while vocalization is a fine motor act. Both can be tested and data registered and analyzed using these same biometrics.

Cognitive loads range from low (using a familiar keyboard or letter board); medium (using an unfamiliar letter board or keyboard but retaining the CRP) and high (combining the unfamiliar letter board or keyboard with an unfamiliar communication partner who offers no physical support). The latter evokes emotional load in ecologically valid situations, like those in real life, when one needs to communicate with different people who may have different social-emotional interaction styles. In addition to these, we included the condition with support from a familiar partner and familiar keyboard, yet requesting the participant and CRP to regress to the type of facilitated communication (FC) support that they practiced in the initial learning stages during MH childhood. Under such conditions, the hand of the partner serves as a supporting anchor to the typer and helps guide the participant's hand as he types. Given the level of independence of MH, we hypothesized that the performance would become more variable than typing the prompted phrase without such support. This regression to childhood style would increase memory loads of latent dyadically shared motor programs. This assay also gave us the opportunity to measure levels of agency and leadership during the performance, as they dynamically change from moment to moment. Furthermore, we hypothesized that spontaneous typing responses to open-ended questions would show a fundamentally different motoric signature from instructed cases, i.e., prompting a phrase selected from those uttered and typed during the open-ended responses phase of the assay. **Table 1** summarizes the cases under study for which we saved video documenting the condition and / or biosensor data co-registering kinematics, HRV and video. **Supplementary Figures 1-2-3** show samples of variations in keyboards and partnering styles.

## 2.6 Spontaneous Motions Beneath Awareness as Motoric Signatures of Authorship

We posit that increasing the level of cognitive load would increase the variability of biomechanical motoric actions and be captured both in the physical movement trajectories in 3D space and in the stochastic trajectories derived from the corresponding micro-variations in linear speed. Prior evidence from the lab supported this hypothesis in neurotypicals [54, 58, 59]. Key to this supposition was the monitoring of spontaneous movements occurring largely beneath awareness. In particular, we monitor the spontaneous motions of the non-dominant side of the body because these segments of the body are out of the intentional control of the participant while performing the biomechanical action [36] and away from the CRP awareness. They escape awareness of the unfamiliar partner and the prompter sitting across the table from the participant as well. Our full focus in fact was on typing by the dominant hand as this was the only typing hand. The incidental motions of the non-dominant side, which spontaneously transpired largely beneath our awareness, were used as motoric signatures indicating the level of authorship in open-ended cases above the levels of cognitive effort that we used throughout the assay.

In a natural setting of assisted communication (e.g., FC, rapid prompting method (RPM), spell to communicate (S2C), etc.) the non-dominant side of the body moves spontaneously out of reach from the CRP/Prompter and even out of reach from the typer while typing. Given our discovery that spontaneous movements are key to our understanding of volition and motor control [16, 36], we precisely focused on the spontaneous, goal-less, incidental movements to reference the levels of cognitive load vs. creative wondering, during typing a phrase to answer open-ended questions and typing that same phrase under instruction. We also monitored it when the CRP assisted the typer by having the typer rest his hand over her hand (see **Supplementary Figures 1-2-3**). We documented video continuously and saved the biosensors' data for all these conditions in **Table 1**.

### 2.7 Kinematics Analyses

The 3D position data from all sensors was buffered and saved every 5 minutes, to avoid data loss during the conversation flow. The Motion Monitor software provides real-time visualization of the upper body positional trajectories, as they unfold (**Supplementary Figure 5**), as well as the ECG data from the EMG sensor, along with a plethora of kinematics parameters. Position data was then analyzed with in-house software. Using MATLAB splines toolbox, we smoothed the positional trajectories and derived the linear speed to examine the peaks and valleys of the continuous time series that they produce. We saved all data and documented the sessions using video and audio recordings. **Supplementary Figure 6** shows the steps to acquire the temporal speed profiles and do detail kinematics analyses.

### 2.8 Stochastic Analyses

The speed profiles were used to examine the peaks and their latencies. The function *findpeaks* from MATLAB was used to obtain the peaks and locations of the peaks (latencies). Inter-peak-intervals (IPI) latencies will be featured in a different paper, due to space constraints. Linear Speed peak amplitudes were gathered into a frequency histogram and the distribution fitting toolbox in MATLAB used to determine the best fitting distribution. Maximum likelihood estimation (MLE) functions were also used to probe different scenarios. The continuous Gamma family of probability distributions was the best fit in an MLE sense. We used 95% confidence intervals to *empirically* determine the Γ shape and scale parameters of the micro-movement spikes (micro-peaks) derived from the linear speed peaks. Then we obtained the Γ moments (mean, variance, skewness and kurtosis) for each set of amplitude and latency parameters. **Supplementary Figure 7** shows the pipeline of these analyses which have been previously published and patented.

### 2.9 Scaling Power Laws of the MMS Micro-peaks and Gamma Parameters

The log-log of the Γ-parameters follows a scaling power law [16, 60]. We obtain the slopes, intercepts and goodness of fit for each condition. We do so as well for the Γ differential entropy as a function of the Γ-shape. **Supplementary Figure 8** shows representative linear fits of three conditions and all values are reported in **Table 2**.

The Γ distribution parameters are locked into a geometric constraint where both variance and differential entropy become tied to a single driving variable, namely the shape $(a)$ .

$\ln(b) = -M\ln(a) + C$ where M > 0, which by exponentiation simplifies to the power law

$$b = e^{C}a^{-M} \qquad \text{(Equation 1)}$$

### 2.10 The Gamma Variance as it Relates to the Scaling Power Law

$$\sigma^2 = a \cdot b^2 = a\left(e^{C}a^{-M}\right)^2 = e^{2C}a^{1-2M} \qquad \text{(Equation 2)}$$

The physical meaning of this equation is that the variance follows a power law driven entirely by the shape parameter $(a)$. In equation 2, if M = 0.5, the exponent becomes 0, locking the variance at a constant value of $e^{2C}$ but if M > 0.5, the variance collapses toward 0, as the shape value grows. In Equation 1, we will obtain the values of the exponent M (the slope) and examine the differences across assays.

### 2.11 The Gamma Differential Entropy as it Relates to the Gamma Shape and Variance

The general definition of the differential entropy of a continuous random variable X is:

$$H(X) = -\int_0^{\infty} f(x)\ln f(x)\,dx \qquad \text{(Equation 3)}$$

For a standard Gamma distribution (with our scaling power law for humans):

$$H(X) = C + a - M\ln(a) + \ln(\Gamma(a)) + (1-a)\Psi(a) \qquad \text{(Equation 4)}$$

So, we express the differential entropy as a function of the Γ shape and the slope M acts as a modulating factor. The Γ shape $(a)$ dictates both equations 3 and 4, the dynamic behaves entirely differently based on the value of the slope *M*. A question therefore is whether there are differences across assay conditions with different cognitive loads capturable by these relations. To address that question, we obtain the values of M (and C the intercept) for each of the 12 log-log shape vs. scale scatters of each assay and do so as well for the differential entropy scatters as a function of the shape. We report on the ranges of M and C in **Supplementary Table 1**.

### 2.12 Possible scenarios for Slope M

Case 1 – High inverse scaling: When M > 0.5 the scale decreases fast as the shape increases. As $a \to \infty$, the variance $\sigma^2 = e^{2C}a^{1-2M}$ shrinks towards 0 $(\sigma^2 \to 0)$ because the exponent $1-2M$ is negative. The entropy drops down toward negative infinity. The severe compression of the scale parameter overpowers the spreading effect of the shape parameter.

Case 2 – Low inverse scaling: When M < 0.5 the scale decreases slowly as the shape increases. As $a \to \infty$, the variance grows toward infinity $(\sigma^2 \to \infty)$ because the exponent $1-2M$ is positive. The entropy grows toward positive infinity. The broadening shape parameter $(a)$ dominates the mild compression of the scale $(b)$.

Case 3 – The case when the critical threshold *M = 0.5*. In this case, $b = e^{C}a^{-0.5} = e^{C}\frac{1}{\sqrt{a}}$. This causes fixed variance at a constant value, completely decoupled from the parameters: $\sigma^2 = a\left(e^{C}\frac{1}{\sqrt{a}}\right)^2 = e^{C}$. The entropy depends purely on the shape since the variance is strictly locked at $\sigma^2 = e^C$, any change in the shape $(a)$ modifies the entropy purely by changing the geometry of the distribution (moving from highly skewed exponential-like profile to a symmetric bell curve) without changing the overall standard deviation. As the shape $(a)$ becomes large, the Gamma distribution approaches the normal distribution.

### 2.13 Agency Index

In prior work, we developed indexes of socio-motor agency as a ratio of motor control to motor autonomy [4]. Motor autonomy is expressed as approximate entropy, a measure of order, ranging from low as completely ordered (deterministic) to high as totally disordered (chaotic). From the standpoint of an external controlling agent, at the chaotic end, the system of the agent being controlled is uncontrollable and as such, it has full autonomy. The other extreme (deterministic) reflects, from the standpoint of the external controlling agent, that the system being controlled is fully controllable and as such, it has no autonomy. In the situation where the CRP guides the hand of the participant, we wanted to quantify the extent to which the CRP was fully surrendering her autonomy (high Approximate entropy) and as such letting the participant drive the movements. This is a situation whereby the CRP motor system guides the other without overpowering control and in the case of being instructed, the participant surrenders his full autonomy to follow instructions.

The other term in the agency ratio refers to predictive motor control. To obtain this index, we relied on the scaling power law characterizing Gamma processes of MMS whereby the Γ-NSR (variance to mean ratio) provides a range including at one extreme predictability and signal certainty and at the other extreme, randomness and memoryless processes characterized by high noise. A system with good predictive motor control has low Γ-NSR, which also corresponds to high values of the Γ-shape, tending towards the Gaussian regimes of the continuous family of Γ-PDFs. Here is the index according to equation 5 below:

$$Agency = \frac{PredictiveControl}{PredictiveControl + Autonomy} = \frac{\Gamma_{SNR}}{\Gamma_{SNR} + MidApproxEntropy} \quad \text{(Equation 5)}$$

The $\Gamma_{SNR}$ is $1/\Gamma_{NSR}$, where NSR is empirically estimated as the scale parameter of the continuous Gamma family of PDFs, because $\Gamma_{NSR} = \frac{\Gamma\sigma^2}{\Gamma\mu} = \frac{a \cdot b^2}{a \cdot b} = b$

This ΓSNR quantity is normalized between 0 and 1 in Equation 6, by obtaining all the values across the participant and the CRP across the assays in the session as:

$$Normalized\ \Gamma_{SNR} = \frac{x - \min \Gamma_{SNR}}{\max \Gamma_{SNR} - \min \Gamma_{SNR}} \quad \text{(Equation 6)}$$

where $x$ is the estimated ΓSNR for angular speed MMS (or acceleration MMS) each assay-segment of the session, and the minimum and maximum are taken across all frames of the session, pooling across the typer and the CRP data sets.

The Approximate Entropy is similarly normalized between 0 and 1 as in Equation 7:

$$NormalizedApproxEntropy = \frac{x - \min ApproxEntropy}{\max ApproxEntropy - \min ApproxEntropy} \quad \text{(Equation 7)}$$

where $x$ is the estimated Approximate Entropy value of each assay in the session, and the minimum and maximum are taken across all frames, pooling across the typer and CRP data sets. Then we obtain the midpoint of the Normalized Approximate Entropy in Equation 8:

$$MidApproxEntropy = \left(\max NormalizedApproxEntropy - \min NormalizedApproxEntropy\right) / 2 \quad \text{(Equation 8)}$$

By tracking this ratio defining socio-motor agency, we seek to understand the moment-by-moment evolution of the interaction between the CRP and typer, particularly during assisted typing (FC).

### 2.14 Transfer Entropy Analyses

We used the Infodynamics-dist toolbox (version 1.6) [61] in MATLAB to compute continuous transfer entropy between the typer's sensors and the CRP sensor. To that end, we used the angular speed from the gyroscope sensor. We obtained pairwise measures of Transfer entropy (TE) to quantify the directed flow of information between the time series of micro-peaks in the MMS obtained from the sensors of the typer and that on the wrist of the supporting (dominant) hand of the CRP. We seek to quantify statistical causality by measuring how much knowing the past of one system (the typer set as the source) reduces uncertainty about predicting future events of the other system (CRP set as the destination), beyond what its own past already tells it. We then invert the source and destination and obtain a matrix of asymmetric entry values indicating (through the rows) which body part from which agent of this dyad, leads the interaction. We look 100ms and 200ms into the past as these are appropriate ranges of latencies for visual and motor-reafference sensory processing [62-66].

### 2.15 Cross-Coherence Analysis

We also computed values of coherence estimates (0-1) as a function of frequency (in 100Hz), obtained between representative angular speed MMS micro-peaks from each assay (ranging between 8,000 and 20,000 frames) across the session. We obtained corresponding cross-spectrum phase (deg) and frequency portrays of each assay. Coherence values above the overall median value across body parts of the typer and dominant hand of the CRP were used to determine degree of (pairwise) synchronization between each sensor on the typer and CRP's supporting hand (via her wrist sensor).

### 2.16 Geometric Analyses Stochastic Space

In our stochastic analyses, the points of our parameter spaces are Γ-probability distribution functions. Since the Γ-parameters (shape and scale) span a curved manifold [67] where Euclidean distance is inappropriate as a similarity metric to measure change across the Γ-PDFs, we adopt the Fisher-Rao (FR) metric [68]. This metric is appropriate to measure the distance between two Γ-PDF points of this space of interest, especially to separate changes in Γ signatures of motoric micro-peaks with increase in cognitive loads [69]. Furthermore, we obtain the geodesic (shortest distance paths) between Γ-parameter pairs representing different Γ-PDFs across the body localizing the condition as points on our parameter space. These geodesic path values of the shape and scale parameters can then be used to obtain the corresponding statistics path spanned by $\Gamma\mu$ and $\Gamma\sigma^2$.

We measure the FR-distance between the Γ-PDFs corresponding to the right and left hand for each condition. Furthermore, we examine representative geodesics for the experimental conditions of interest. These conditions are presented in Table 1.

Once we plot the right-hand vs. left-hand FR-distance values across conditions, we obtain representative points to plot and contrast the corresponding geodesic paths between the right dominant hand and the left non-dominant hand. We also obtain the pairwise FR-distance from each sensor location (body parts in **Figure 2A**) to each one of the other body parts. This symmetric matrix of FR-distances is displayed as a color map with blue hues representing high similarity (small FR-distances) and yellow orange hues representing low similarity (large FR-distance) on the Γ-manifold.

### 2.17 Geometric Analyses Physical Space

We analyzed the geometry of the physical trajectories in 3D space. This was important to establish correspondence between the stochastic and physical spaces and to investigate the kinematic parameters and the Γ-parameters that best represent the different levels of cognitive load, in the order in which they evolved over the session. Physical distance analysis anchors the abstract stochastic manifold analysis to our physical reality.

First, we used the speed profile to determine the curve's length (for each sensor on the right and left hand) by cumulatively adding the speed magnitude along the path. We use the differential geometric definition of path length (arc length) $s$ of a continuously differentiable curve $\gamma:[a,b]\to M$ on a Riemannian manifold $M$, defined as the integral of the speed of the curve with respect to the metric tensor in Equation 9 [70]:

$$s=\int_a^b\left\|\frac{d\gamma}{dt}\right\|dt=\int_a^b\sqrt{g_{\gamma(t)}\left(\frac{d\gamma}{dt},\frac{d\gamma}{dt}\right)}dt \quad \text{(Equation 9)}$$

Where $\gamma(t)$ is a smooth curve mapping the interval $[a,b]$ into the manifold $M$, $\frac{d\gamma}{dt}$ is the tangent vector to the curve at parameter $t$, residing in the tangent space $T_{\gamma(t)}M$ , $g$ is the Riemannian metric tensor, which defines the inner product on each tangent space and $\|.\|$ is the norm induced by the metric, where $\|\upsilon\|=\sqrt{g(\upsilon,\upsilon)}$

Second, we obtained the absolute difference between the physical distance computed from the dominant hand vs. the path length of the non-dominant hand speed trajectory. Lastly, we plot them in the same format as the stochastic FR-distance of the corresponding empirically estimated Γ-PDFs for the two hands. We asked if there was congruence in the trends of the FR-distance on the abstract stochastic manifold description of the motoric signatures and the distance derived from the 3D physical trajectories. Notice that we employed the Euclidean line element to obtain the speed magnitude of the 3D velocity field (the local tangent vectors to the curve at each 3D positional point) obtained by properly smoothing and differentiating the 3D-positional trajectories of the hands (and the other sensors). When obtaining the path length, we used the cumulative sum of the speed magnitudes in Equation 9. As such, our calculations reflect the true pathlength of the curved trajectories. They are not the Euclidean distance between the initial position $x_{init}$ of the hand and $x_{final}$ wherever the hand ends when the condition is finalized.

# 3 Results

## 3.1 Non-Dominant Body Physical Movement Trajectories Automatically Differentiate Independent Spontaneous, Instructed and Guided Performance

### 3.1.1 Warming up conversation

We recorded spontaneous, self-initiated conversations between the participant and the experimenters (Saved video and biosensor): These consisted of consenting to the study, answering various open-ended questions and questions that required a Yes, No or Maybe answer. They included discussions about different aspects of the study, comfort levels about the sensors and instruments, comfort level with the cameras filming everything we were about to do and various people in the lab coming and going in the background as if in a natural environment. This was important to lend the study ecological value. **Figure 4** shows the trajectories and stochastic analyses from the micro-peaks of the standardized micro-movement spikes (MMS) derived from the temporal speed profiles (signal amplitude) of the sensors' trajectories. The 3D trajectories of the center of mass of each body part (depicted in **Figure 2A**) are accompanied by the corresponding temporal speed profile in **Figure 4A**. The other panels show the empirically estimated ΓPDF from each sensor's micro-peaks of the linear speed's amplitudes along the time series.

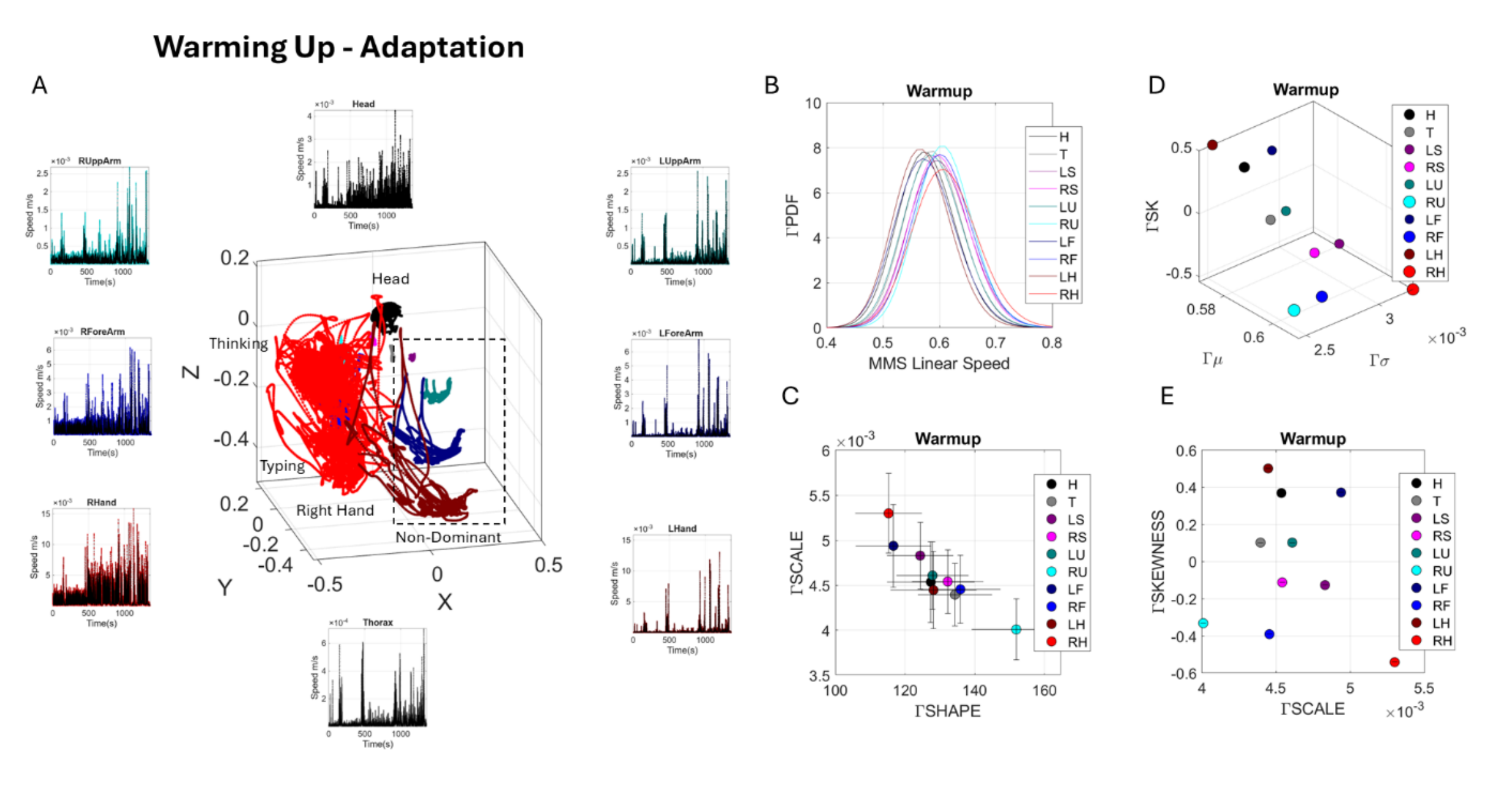

**Figure 4. Warming up and adapting to the lab environment**. (A) The 3D physical trajectories and corresponding temporal speed profiles from each body part, including the dominant right side and the non-dominant left side. (B) Empirically determined Γ probability distribution functions (PDFs). (C) Empirically determined Γ parameters with 95% confidence intervals for each body part. (D) Corresponding Γmoments. (E) Projection of the Γ scale (Noise to Signal Ratio) vs. Γ Skewness.

### 3.1.2 Spontaneous, open-ended communication with haptic support

We used open-ended questions to evoke conversation: Saved video and biosensor,

EBT. What do you think about today's meeting here?

MH: "*I think that the degree of communication that is not spoken affects the quality of the relationship, I mean you have great non-verbal communication between Daniel and Liz.*"

This was in reference to another researcher (Dr. Daniel Orlievsky, to the right of MH in **Figure 1**) who was present during the session and communicated with EBT gesturally as we observed the participant perform the tasks. MH perceived subtle facial exchange and verbalized it both through the spelling on the keyboard and by vocalizing the phrase.

MH: *"I want to find a way to bring typing the words to many because I know that silence to a person is like you live in your own parallel universe behind glass".*

EBT: Tell me, do you remember way back, when you were little, and you started typing, do you remember what you felt when you started typing and saying things that you could, that people reacted to you and spoke back to you? And then there was a conversation going on? Do you remember? That was a long time ago, probably … do you remember, because that must have been extraordinary, when you started to get a reaction from people.

MH: *"I think you can have typing but still be silent because the typing must express the big feelings inside. I was always thinking big things so typing the words was like appearing in the sky myself visible."*

**Figure 5** shows sample trajectories from open-ended communication with haptic support consisting of light touch of the right knee to calm the body. The trajectories of the dominant side were then more economical than the adaptation case (shorter 3D excursions), but the non-dominant side remained engaged. This is appreciated in the amplitude of the linear speed profiles and in the stochastic Gamma signatures. Transfer of motor noise from the dominant typing side alone could not explain this differentiable engagement of the non-dominant side.

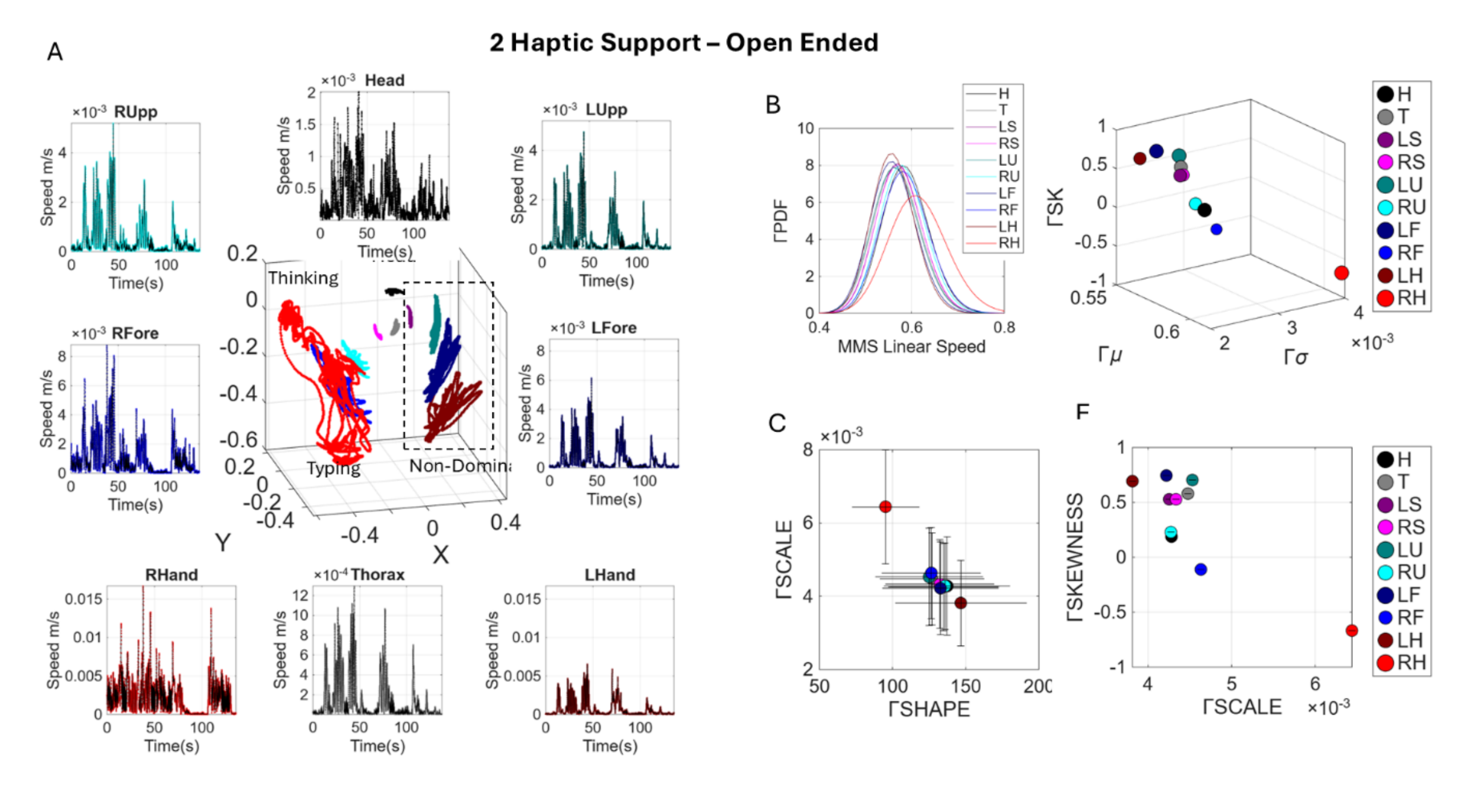

**Figure 5. Typing answers to open-ended questions with light haptic support on the right knee to help control and dampen right legs' motions**. (A) 3D positional trajectories of dominant and non-dominant sides and corresponding linear speed temporal profiles. (B) Empirically estimated Γ PDFs. (C) Γ parameters with 95% confidence intervals. (D) Γ moments. (E) Projection of Γ skewness and Γ scale (NSR).

### 3.1.3 Spontaneous, open-ended communication without haptic support

**Figure 6** shows the case when no support was provided during part of the communication exchange. Saved video and biosensor: The non-dominant side of the body remained engaged, as shown by corresponding speed profiles and Gamma stochastic profiles. As in the supported case of **Figure 5**, the non-dominant side of the body presented modulated responses with the thinking process that took place while elaborating the answers to the open-ended questions. This thought process posed higher cognitive load which funneled out into the motoric code. How to separate in the stochastic and physical signatures these changes in cognitive load then require a radically different version of the task that quenches the variability to the intentional effector, i.e., the right hand typing the phrases. To that end, we next try the instructed version of the task whereby we choose a phrase from the answers to open-ended questions and repeatedly prompt MH to type it under different levels of cognitive and memory loads.

Additional sample open-ended questions and spontaneous answers transcribed from the videos are below.

EBT: Was there a moment that you felt that changed, for you? That you see that people are seeing it? Was there a moment that you felt that way or is it still that you feel invisible?

MH: *"I am sure that my experience is not typical because I was surrounded by people who believed in me and people who did not, I have the compassion for."*

EBT: Do you communicate with other typers, do they feel the same way, that is, they are supported by people who believe in them but are also aware of those who don't? Other typers that may be your friends?

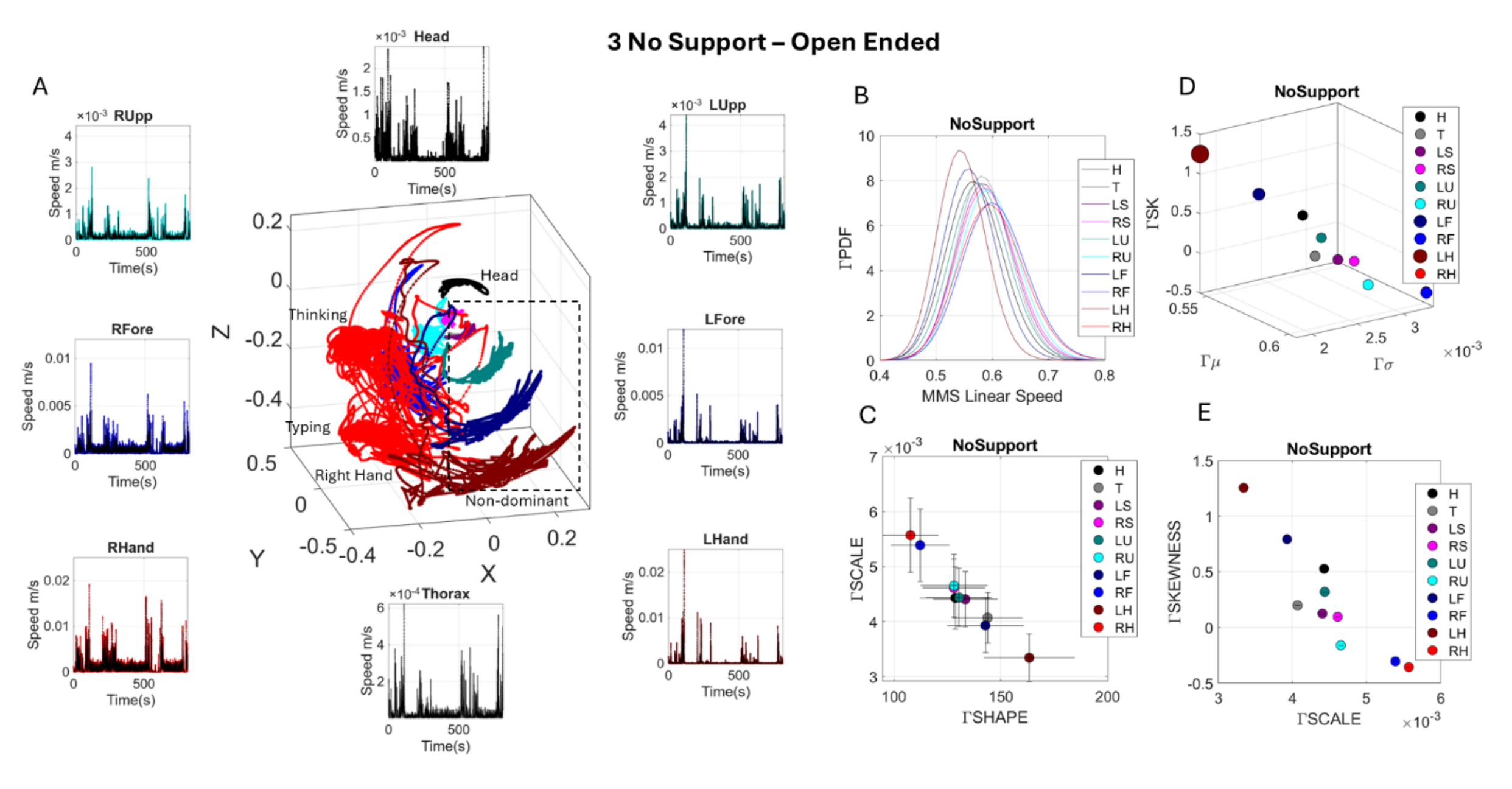

**Figure 6. Motoric trajectories in 3D space from the dominant and non-dominant body parts during the spontaneous typing of answers to open-ended questions without any support**. (A) 3D positional trajectories and corresponding linear speed profiles. (B) Empirically estimated Γ PDFs. (C) Empirically estimated Γparameter scatter with 95% confidence intervals on the Γ parameter plane. (D) Corresponding Γ moments. (E) Projection of Γ skewness vs. Γ NSR.

MH: *"I think that the typers are not always feeling so supported so the voices of doubt cut more deeply the greater the lessons of discouragement do to them".*

***"I am going to tell you that millions are pinning their hopes on our great work that is creating a powerful force of forward momentum."***

I chose this phrase to test the instructed communication condition. Instructed communication has fundamentally different demands in thought preparation than open-ended typing. There is no new thought to produce the prompted phrase. Biomechanically, the motoric demands of instructed communication of a phrase should be like those of the previously typed phrase in spontaneous open-ended mode. **Figure 7** compares the two cases side by side, using the same phrase.

### 3.1.4 Instructed communications of prompted phrase using a familiar keyboard

The participant performed the biomechanical task of typing the exact same prompted phrase three times. Saved video and biosensor: The results were consistent as in representative trial **Figure 7**, right panel. The non-dominant side of the body became dormant. The speed profiles dampened the amplitude relative to the dominant side typing the instructed phrase. Furthermore, the changes in the physical 3D space manifested as well in the Gamma signatures. These are shown side by side on the bottom panels of **Figure 7** for the Γ moments space (left) and the Γ parameters' plane (right).

The Γ skewness profile of the non-dominant side of the body dramatically shifted from spontaneous thoughts expressed in response to the open-ended question, to the same prompted phrase deliberately typed under instruction. In the former case, the non-dominant side had heavy tailed distributions to the right (increasing above 0 value which is the symmetric range) while in the latter, the fluctuations shift the heavy tail to the left (smaller fluctuations).

Heavier right tailed Gamma distributions indicate accumulations of larger fluctuations in MMS micro-peaks amplitudes. According to the Equation 10 ratio, larger numerator values signify that on average, the deviations of the raw speed amplitudes from the empirical $\Gamma\mu$ are smaller. The denominator does not penalize the numerator. In contrast, smaller fluctuations in MMS micro-peaks amplitudes indicate that on average the fluctuations from the $\Gamma\mu$ are large. The denominator penalizes the numerator. Equation 10 below, scales out allometric effects due to anatomical disparities across the population [71]. Supplementary Figure 7A shows the absolute deviations from the empirical $\Gamma\mu$ of a sample instructed trajectory. These deviations are subject to Equation 10 and then the full MMS micro-peaks in Supplementary Figure 7B used in the estimation of the Gamma parameters, geodesic distances, geodesic paths and corresponding moments).

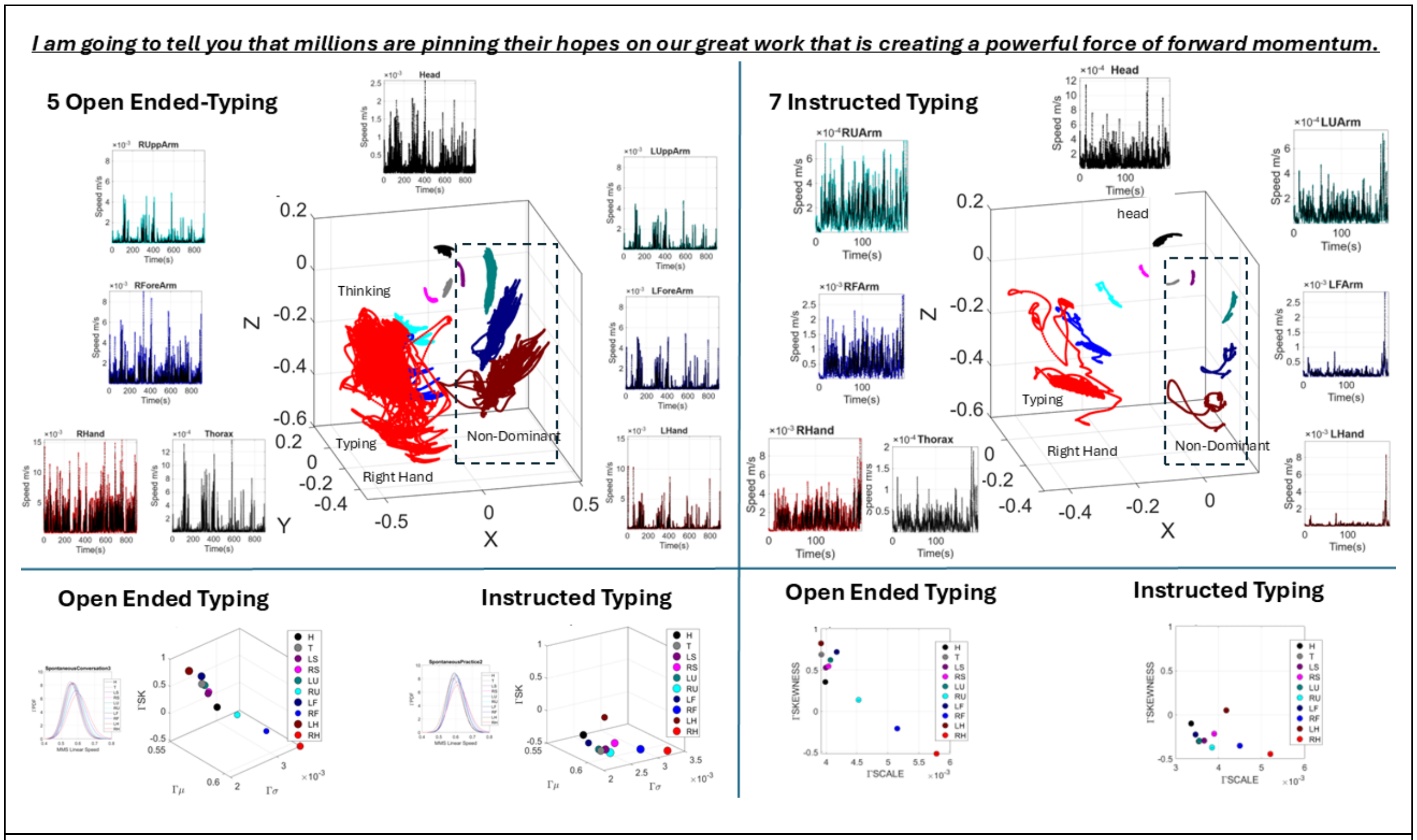

**Figure 7. Motoric signature of communication authorship during typing (and/or spoken phrases).** Comparison of non-dominant bodily trajectories, corresponding speed profiles and empirically estimated Gamma stochastic signatures when spontaneously typing the phrase above in response to the open-ended question (see main text) vs. the same phrase prompted in the deliberately instructed condition. Note the dramatic shift in both physical and corresponding stochastic motoric signatures of the non-dominant side – entirely beneath awareness of the typer and CRP. The CRP provided slight haptic support to the right (dominant) knee to help dampen involuntary motions.

$$MMS = \frac{Peak}{Peak + Avrg_{\min to \min}} \quad \text{(Equation 10)}$$

Here the *Peak* refers to the local speed amplitude (local maxima in the time series of peaks and valleys) and $Avrg_{\min\ to\ \min}$ refers to the averaged speed values obtained between the two local minima surrounding the local maxima (Peak). As the average speed tends to decrease in more controlled movements towards the targeted letter on the keyboard, the numerator grows larger. Then the MMS micro-peaks fluctuate more around the $\Gamma\mu$ of the raw speed peaks. These larger MMS fluctuations accumulate to the right and explain the shift to the heavier right-tailed Γ-distributions range. The dominant hand and forearm intentionally typing the phrase are not impacted. In contrast, when the speed increases on average in less cognitively controlled, more automated motions, the MMS micro-peaks fluctuate less. The overall ratio yields smaller MMS micro-peaks' values. Such small fluctuations accumulate to the left and explain the shift in skewness to the heavier left-tailed Γ-distributions range.

Notice that regardless of the spontaneous typing of the phrase from open ended questions or deliberately typing the same phrase under prompted instruction, the typing (dominant) right hand and right forearm remain in similar Γ-regime. It is the non-dominant side of the body which shifts stochastically, consistently capturing the changes quantified in the 3D physical trajectories. The non-dominant side of the body is largely beneath awareness, yet it funnels motorically the changes in cognitive load. Notice as well that fatigue alone cannot account for such changes because we bring them back down to a dormant state with instructed prompted phrases and systematically re-engaged them with higher cognitive, memory and emotional loads.

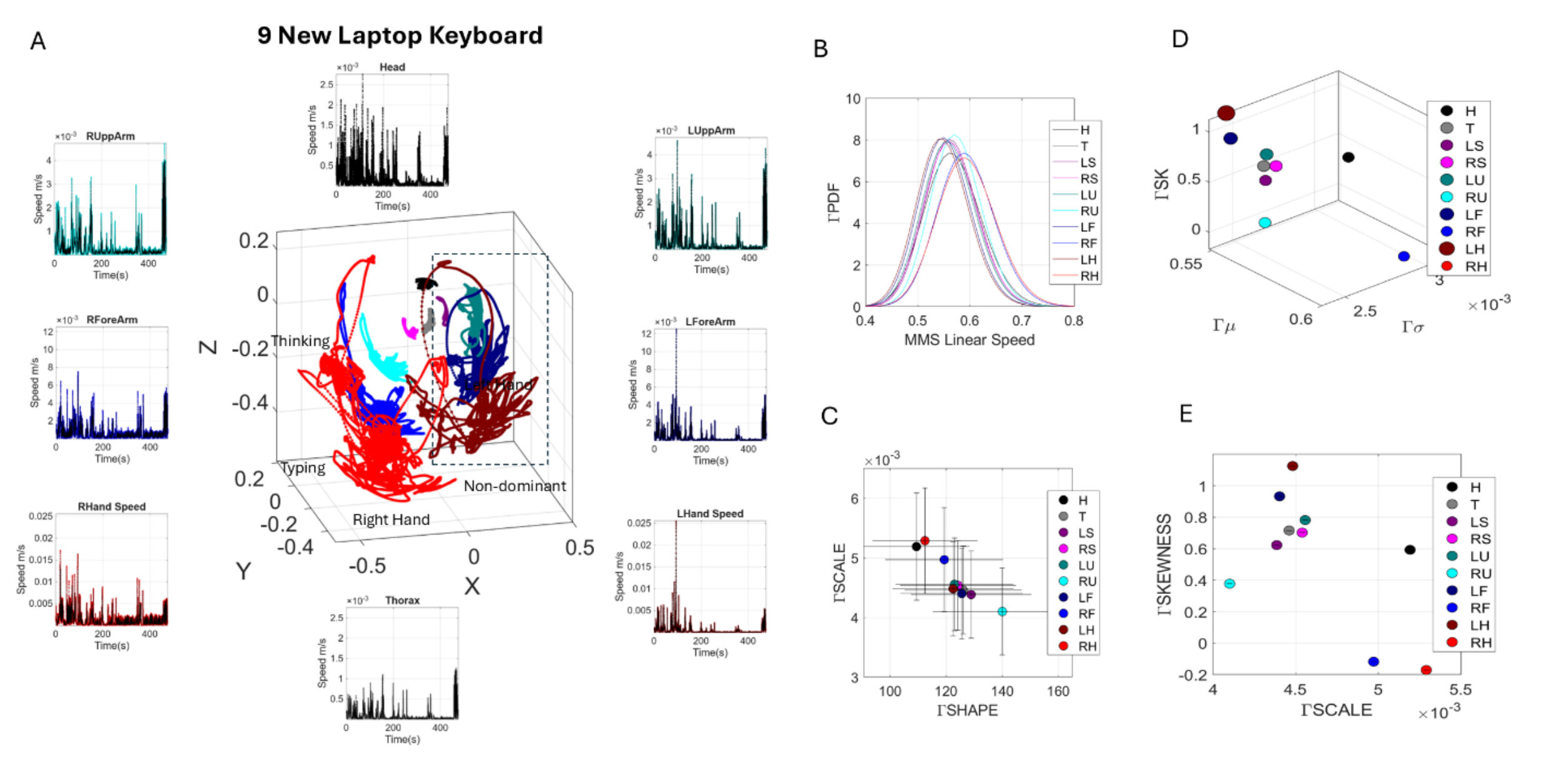

**Figure 8. Instructed prompted condition (high cognitive load) using an unfamiliar keyboard, unfamiliar communication partner and different prompting style** re-engages the non-dominant body part during instructed spelling of words e.g., 'complex'. All panels follow the same format as before.

### 3.1.5 Typed phrases using a different keyboard layout with an unfamiliar partner

Prompting and instructed typing were performed in other contexts under higher cognitive load (as in **Table 1**). Saved video and biosensor: We manipulated the degree of novelty of the environment (familiar or unfamiliar letter board, keyboard, partner, fully supported hand by familiar partner). Below we explain the rationale behind each task in the instructed component of the assay.

This condition switched the familiar keyboard to the keyboard embedded in a laptop with different layout, completely new to the participant. This required a mental remapping of the keys, the orientation tilt of the letters, the visual contrast of letters foreground and background and overall distances between letters. Furthermore, the communication partner changed to an unfamiliar person who provided no haptic support throughout the exchange and had a different prompting style than EBT.

This high-cognitive-load condition required spontaneous thinking to remap the keyboard configuration and type the words on command. Furthermore, the context changed. With the unfamiliar partner and different levels of emotional support, the participant had to remap and recalibrate his motor performance. Owing to all the cognitive and emotional (uncertainty) loads that this condition brought, the assay required a geometric transformation localizing the new keys layout, planning a different motor trajectory to a different target (letter) and executing the goal-directed action, all without the familiar CRP. This higher cognitive load case is depicted in **Figure 8** showing the motor output of the spelling-words version of this assay. **Figure 9** shows the motor output of the prompted phrase version of the assay, under deliberate instruction**.**

As in **Figure 7** above (right panel Instructed condition) the stochastic Gamma signatures of the non-dominant side of the body shifted skewness upwards (like in spontaneous open-ended typing), except for the dominant right hand which remained impervious to the manipulation in cognitive load. This is expected as in both situations the hand is deliberately delivering the typing strikes to the intended targeted letters on the keyboard. Similar patterns emerged in the instructed case of **Figure 9** when the full phrase was prompted. The non-dominant side of the body re-engaged under higher cognitive load.

### 3.1.6 Instructed communications using familiar letter cardboard

MH uses a familiar yellow cardboard with black letters for "one-finger-typing" (see **Supplementary Figure 3 top-right panel**). Saved video only**:** In this very familiar configuration, MH used single strikes on the first letter of a whole phrase and would then say the phrase. Unlike the familiar or unfamiliar keyboards, where he typed all the letters of the words being typed, here it sufficed to type fewer letters of the phrases being spoken. This board provided a much more economical arm-hand motor program synergy to evoke the spoken phrases. In the language of Differential Geometry, a single strike could be represented by a geodesic direction spanning the full (unique shortest distance) geodesic path to the target [24]. This is an element of the tangent bundle encompassing a family of such geodesic directions, representing geodesic trajectories of entire motor programs. MH figured out how to compress all that information and link a single strike to a full phrase. Using our Γ-space, it is now possible to capture such economical representation geometrically to accelerate the learning process of novice typers.

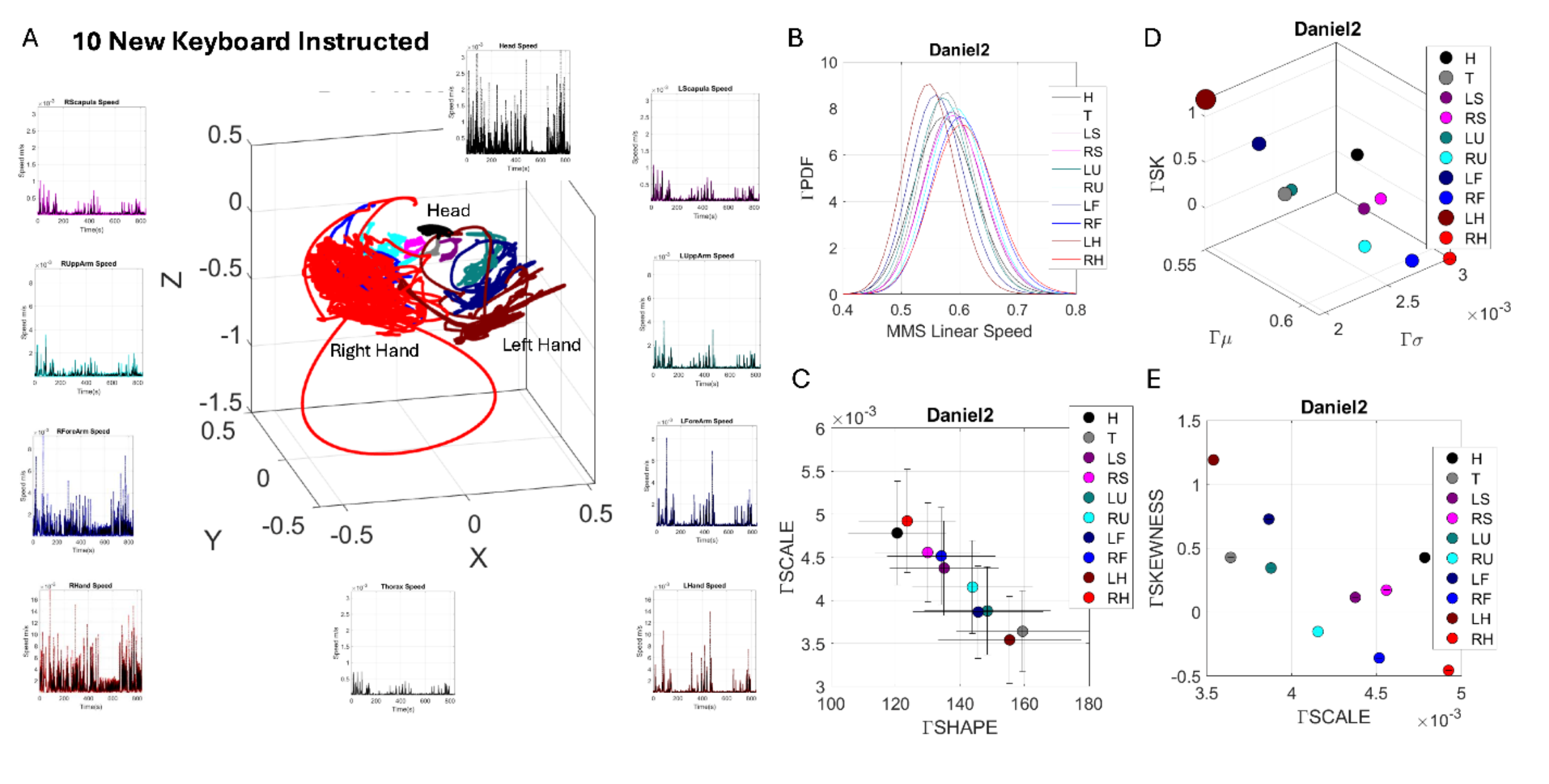

Figure 9. **Instructed with high cognitive load**. Unfamiliar keyboard, unfamiliar partner, different prompting style and same instructed phrase. All panels follow similar formats as before.

### 3.1.7 CRP-Guided communications using the familiar cardboard evoking economical motor programs to communicate, but this time covered with a white piece of paper, so the letters were not visible

In this condition, we increased the cognitive load of the task by covering the keyboard with a white piece of paper such that the letters were not visible to the participant. Saved video and biosensor: To type on the keyboard and evoke the phrases articulated by his mouth, there was a required memory recall of the keyboard layout and a remapping of the motor program. Furthermore, this condition required FC or guided typing (See **Supplementary Figure 3**). Under guided typing the typer's hand rests over the CRP hand. This condition was further difficult because it required regressing to the childhood learning, under supported typing, whereas at present MH's typing is independent. We hypothesized that this type of motor program recall would evoke higher variability along a distinct geodesic path. We further used the condition to explore agency and leading patterns through TE analyses in the next section.

**Figure 10** depicts the motor output and stochastic signatures of one trial of this condition, which we repeated twice. As the previous instructed conditions with higher cognitive load the Γ skewness increased (heavier right-tailed Γdistributions) across the non-dominant (left) side of the body, while the dominant (right) hand and forearm remained impervious to the manipulations. The deliberate movements of the right hand to the targeted letter remained on the intended course while the unintended motions of the left-hand side funneled out, through the motoric fluctuations, the increasing cognitive load.

### 3.1.8 Instructed Facilitated Communications using full support by familiar CRP

As explained, considering the level of independence at typing and the level of automatic retrieval of pointing motor programs in tandem with spoken words, I requested the participant and CRP to regress to a form of facilitated communication FC during typing. This type of guided typing is commonly used in the initial stages of learning to type. We hypothesized that the motor variability captured along the positional trajectories would be impacted. We received support for this supposition in **Figure 10**.

**Figure 10** contains the positional trajectories of both the FC and the covered keyboard assay. Here we further used the gyroscopic data featuring minute rotations of the CRP sensor on her wrist and of all the typer's sensors across the upper body. We first obtained measures of cross-coherence, a frequency-domain statistic used to quantify the linear relationship, correlation strength, and phase stability between two signals (CRP and typer) across different frequencies.

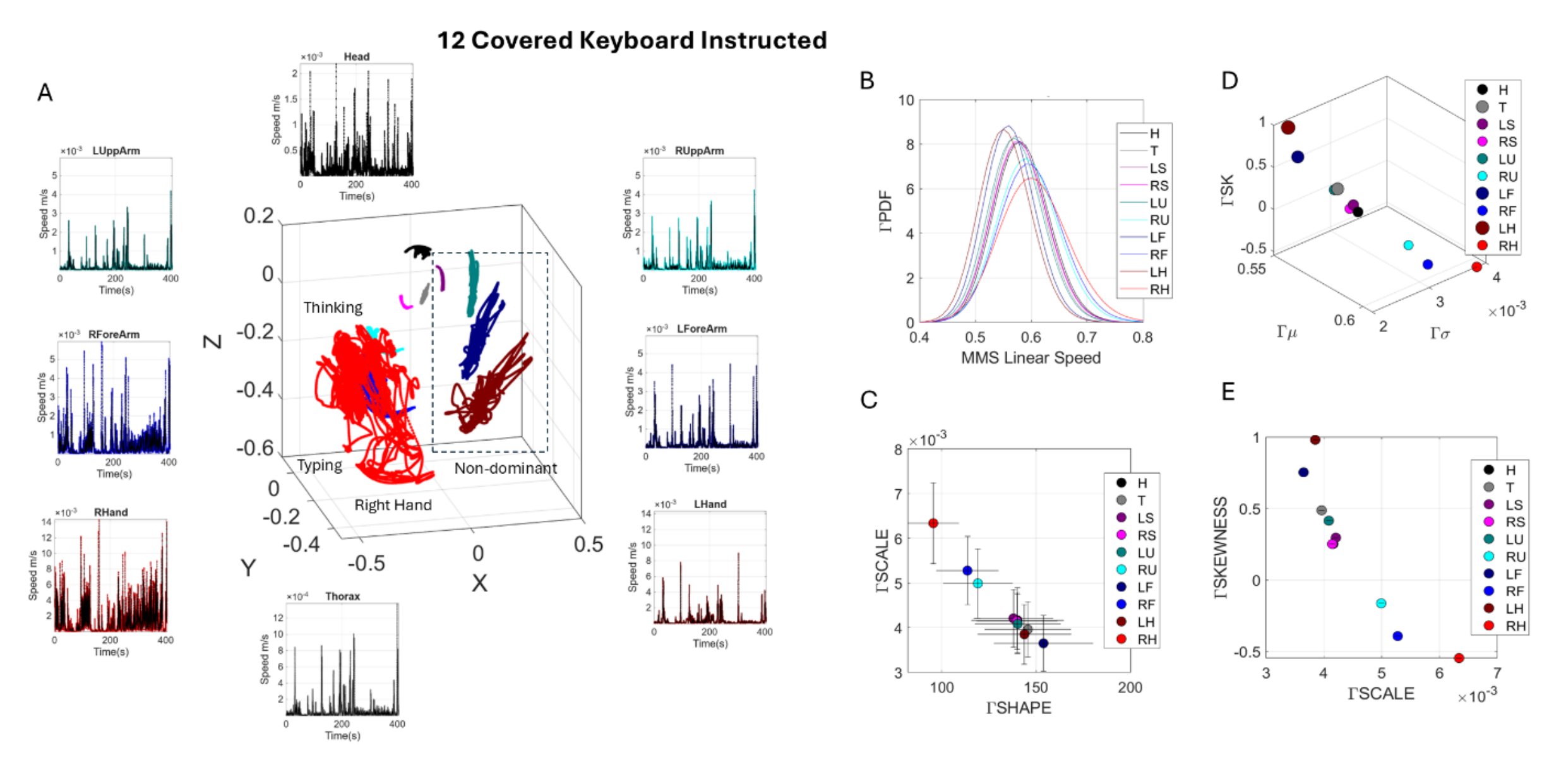

**Figure 10. Same instructed phrase with CRP FC assistance and covering the familiar pocket-size letter board** with a white piece of paper to block the letters and evoke the need to retrieve a mental map of the keyboard to type the start of the words in a single arm-movement strike that evoked the vocalization of the words. Notice re-engagement of the non-dominant side of the body with the increase in cognitive and memory loads. Two partners were present as in Figure 1, one familiar CRP and one new who uses no haptic support.

**Supplementary Materials Figure 11** shows the cross-coherence matrix representing pairwise comparisons between each sensor of the typer and the wrist sensor of the CRP. Entries in this matrix are the maximum values of the cross-coherence and next to it is the corresponding frequency matrix representing the frequencies at which the maximum values occur. During the 2 FC trials with covered keyboard, the maximum cross-coherence values between the typer's dominant hand resting on the CRP hand were 0. This indicates the two signals are completely uncorrelated or independent across those frequencies. Due to the non-stationary nature of these signals, we also used the wavelet coherence function in MATLAB and compared the right vs. left hand of the typer as well as the right hand of the typer vs the right wrist of the CRP and the head and thorax of the typer vs. the right wrist of the CRP providing support for the typer's dominant hand. The results are shown in **Supplementary Figure 12**. They capture shifting coherence over time primarily in the low frequency ranges.

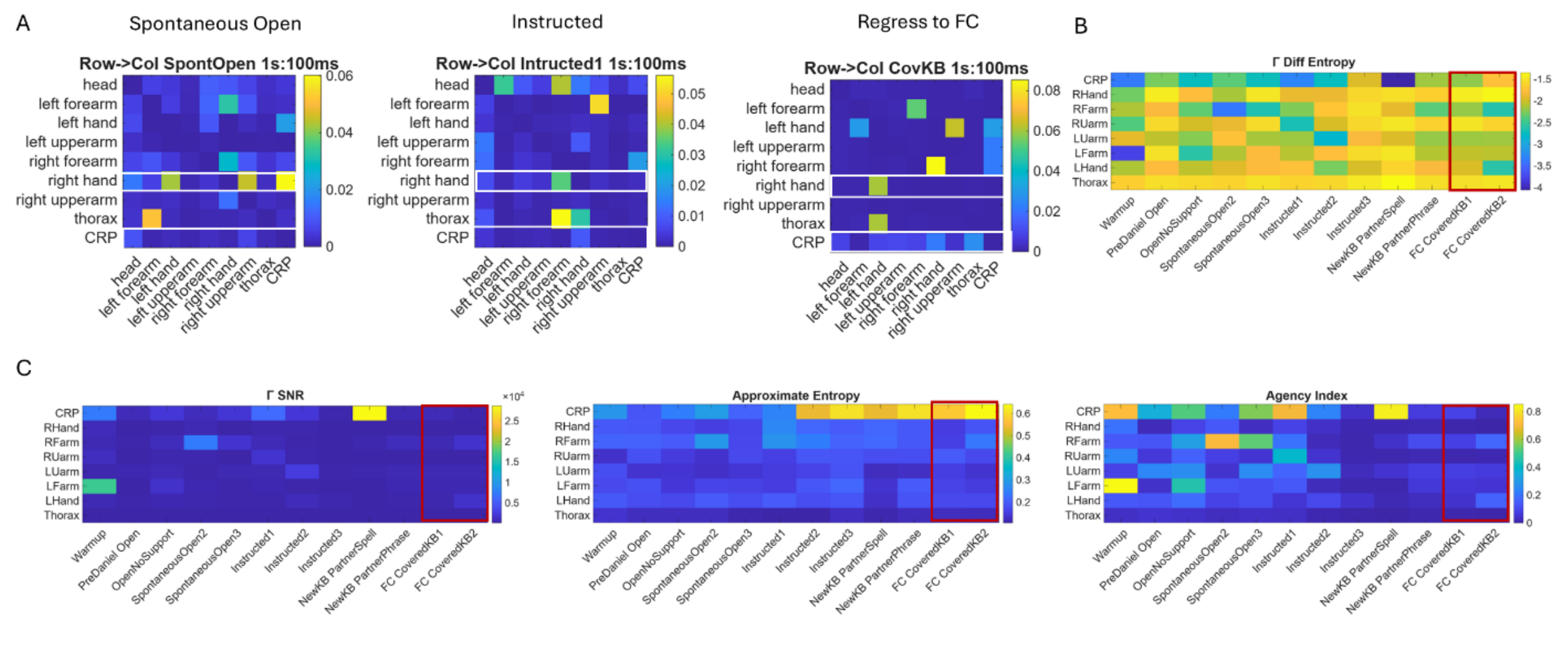

**Figure 11: Transfer Entropy and Agency Index across tasks. (A) Three representative tasks: spontaneous and instructed regression to FC (where the CRP guides the participant's typing hand). The micro-movement spikes (micro-peaks) from the angular speed traces from the gyroscopes were used to obtain the Transfer entropy metric revealing lead of the typer's dominant hand over the CRP hand. (B) Γ Differential Entropy values across tasks and body parts. (C) Agency Index computed as a ratio of Γ SNR and Approximate Entropy obtained for each task and body part. Notice the high Approximate Entropy values of the CRP (representing predictive, deterministic at low and unpredictive chaotic at high values) in regression to FC assay. Red rectangle highlights FC with covered keyboard trials.**

We then examined the transfer entropy values using the CRP wrist angular speed stream as well and the typer's angular speed streams across the sensors on the upper body. We found evidence for the typer's signal leading the CRP at 100ms and 200ms. That is, for those lags, knowledge of the typer's signals lowers the uncertainty of predicting the CRP signals more than just having the CRP signal alone 100ms and 200ms in the past -within latency limits reported for the neural foundations of vestibular, visual and proprioceptive sensory integration signals into the motor streams [63-66].

**Figure 11A** shows this pattern consistently for the 3 main tasks (spontaneous typing to answer open-ended questions, instructed typing without support and FC typing with haptic support and covered keyboard). In all cases the typer's hand (statistically) causally leads the CRP wrist signal. **Figure 11 B and C** shows the computations of the Γ differential entropy values, the Γ-SNR and the Approximate entropy. We used equations 5-8 to obtain a normalized index of agency as a ratio involving the Γ-SNR and the Approximate entropy. **Figure 11 C** shows the matrices thus obtained. **Supplementary Figure 13** shows trends of agency index as a function of Γ differential entropy across bodily sensors while **Supplementary Figure 14** unfolds them by task and sensor location.

We used the MATLAB routines for phase space reconstruction to obtain the time lag (2) and embedding dimensions (3) of the time series (uniformly sampled in time) and returned the reconstructed phase space. We then obtained the approximate entropy, which estimates the likelihood that similar patterns of observations in a time series will not be followed by similar patterns in subsequent increments. Lower values indicate a repetitive, predictable, or regular pattern (deterministic state). Higher values indicate a high degree of randomness, irregularity, or unpredictability (chaotic state).

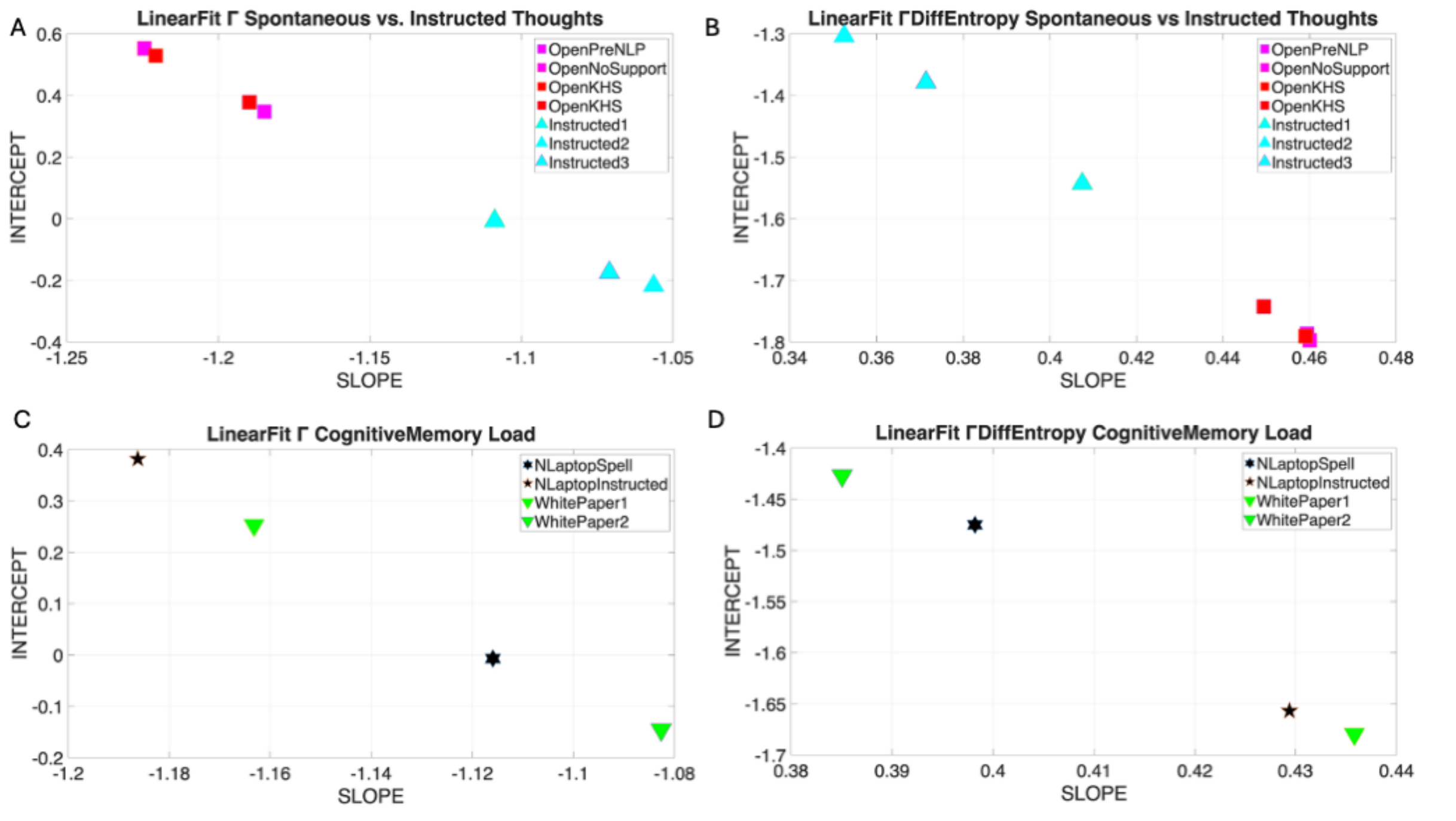

**Figure 12. Motoric signatures of communication authorship from typing. (A) Linear fit to log-log empirically estimated Gamma parameters across the body parts. Red and magenta squares are spontaneously typed phrases (and spoken) in response to open-ended questions while cyan triangles are same phrases but deliberately typed (spoken) by instruction. (B) Similar representation for the slopes vs. intercepts of the linear fit to the empirically estimated log-log Γ differential entropy vs. Γ shape. (C) Cognitive loads during instructed condition. Notice broad range between Γ parameters' signatures and (D) Γ differential entropy.**

In the case of the CRP, the highest values were observed during the FC condition (**Figure 11C** middle panel). This suggests unpredictable patterns of the CRP's wrist when the typer was resting his hand on the CRP's hand while typing. They account for the low agency index values of the CRP during the FC task (**Figure 11C** last panel) and complement the ranges of Γ differential entropy (**Figure 11B**) positively correlated with the Γ-shape (see equation 4). **Supplementary Figure 14** top panel unfolds Agency vs. Γ differential entropy for the CRP right (dominant) wrist supporting the typer's right (dominant) hand as he types in the FC condition. Notice there the low Agency values of the CRP for the FC task. In contrast the bottom panel shows the higher agency values for the typer's sensor on the forearm (leading the hand) during FC with covered keyboard and the highest agency values for the spontaneous typing of his answers to the open-ended questions. These decrease for instructed under higher cognitive load. Unambiguously, the sensor of the forearm broadcasts the typer's agency across all assays. We next describe the results from those stochastic analyses explaining these relations and parameter ranges as rules that enable us to simplify all these computations and reduce them to a handful of parameters whereby knowing one, we can infer the other with high certainty and reduce the computational complexity.

### 3.2 Gamma Stochastic Characterization of Motions in Each Condition Automatically Separates Cognitive Load in Spontaneous and Instructed Assays

Across conditions and body parts, the continuous Γ family of probability distribution functions was the best fit (in an MLE sense) to the MMS (micro-peaks) of the linear speed temporal profiles derived from the 3D physical positional data. The Γ shape and scale parameters were used to characterize each condition as a continuous stochastic process. Their locations on the log-log Γ parameter plane revealed linear relations found also in the Γ differential entropy as a function of the Γ shape parameter. Supplementary Table 1 reports the fitting slope and intercepts and the goodness-of-fit values for both linear regressions across the 12 cases.

### 3.3 Mental Processes Funneled Out Through Motoric Signatures Differentiate Spontaneously Self-Initiated Phrases Answering Open-Ended Questions from Biomechanical Execution of Same Prompted Phrases Under Instruction

**Figure 12A-B** shows the automatic differentiation between the communication through typing of the answer to open ended questions vs. the typing of the same phrase under instruction. This is the case of typing on a familiar keyboard and accompanied by a familiar CRP (see corresponding representative physical movement trajectories in **Figures 5, 6 and 7-**left panel for spontaneous typing to answer open ended questions vs. instructed typing on representative physical movement trajectories of **Figure 7** right panel. Other physical movement trajectories data were used in spontaneous and instructed cases, totaling 96 Γ PDFs, 12 conditions x 8 sensors).

### 3.4 Cognitive, Emotional and Memory Loads Re-Engaged the non-Dominant Side of the Body (Dormant During Instructed Prompted Phrases)

**Figure 12 panels C-D** show the automatic differentiation between cases where the cognitive, emotional and memory loads are higher. The inverted green triangles represent the FC case where a familiar pocket size letterboard was covered by a white piece of paper completely occluding the letters. Across levels of cognitive loads the control and coordination of all the intrinsic rotations of the joints to translate and rotate the end effectors (the hands) in extrinsic space showed smooth movement trajectories accompanied by spoken phrases.

The motion trajectories of the non-dominant side of the body funneled out to the hand motions the mental thought process, preamble to the onset of typing the phrase in response to the open-ended question. When the same phrase was instructed -all things being equal (same partner, familiar keyboard, same table-chair layout to type) the non-dominant side of the body became dormant (see movement trajectories in **Figure 7** right panel). These physical differences in biomechanical movement trajectories captured by the biosensors in physical space transferred to the abstract Γ manifold.

The slopes vs. intercepts of the log-log Γ scale (NSR) vs. Γ Shape and of the log-log Γ Differential entropy vs. Γ Shape revealed unambiguous separation of these spontaneous open-ended vs. instructed (prompted) phrases.

These stochastic characterizations of the positional data featuring the linear speed profiles were congruent with the analyses of cross-coherence, agency, approximate entropy, transfer entropy and differential entropy of the angular speeds reflecting the rotations registered by the gyroscope. The **Supplementary Figure 14** shows the systematic differences during FC (lower agency of the CRP wrist and higher agency of the typer's forearm). Likewise, during spontaneous open ended (independent) typing higher agency of the typer contrasting with his lower agency during instructed and even lower under cognitive loads.

Variations in cognitive loads in Table 1 differentially impacted the physical trajectories of the body. Higher cognitive load led to the spontaneous re-engagement of the non-dominant side of the body after it had become dormant during instructed typing with low cognitive load. Specifically, as in spontaneous *vs*. instructed phrases, here under higher cognitive and memory loads, the slope and intercept planes of the Γ-PDF and Γ Differential entropy parameters automatically revealed the dramatic physical differences.

### 3.5 Gamma Manifold Fisher-Rao Metric and Geodesic Paths Separate Spontaneous Open-Ended from Instructed Motoric Signatures

We explored the Gamma parameters on the curved manifold that they span. To that end, we used the Fisher-Rao distance metric and tracked the distance between the Gamma parameters of the PDFs empirically fitting the dominant vs. non-dominant hands.

This distance was obtained for all 12 cases, in the order in which the conditions occurred during the session. **Figure 13A** shows the progression of the FR-distance between the right (dominant) and the left (no-dominant) hand for the experimental conditions (minus the warming block). **Figure 13B** left shows the geodesic paths corresponding to the longest (spontaneous phrases to open-ended questions) and shortest (instructed phrases) distance traveled on the Gamma manifold, with respect to identical times allotted to both analyses. Right panel shows the geodesics corresponding to the higher cognitive loads under instructed typing. The covered keyboard condition (green geodesics) had congruent paths over the 2 repetitions of the phrase. **Figure 13C** captures the physical distance between the curved path lengths traversed by the dominant and non-dominant hand. Their rate of change trends similarly to the FR-distance on the Gamma stochastic realm.

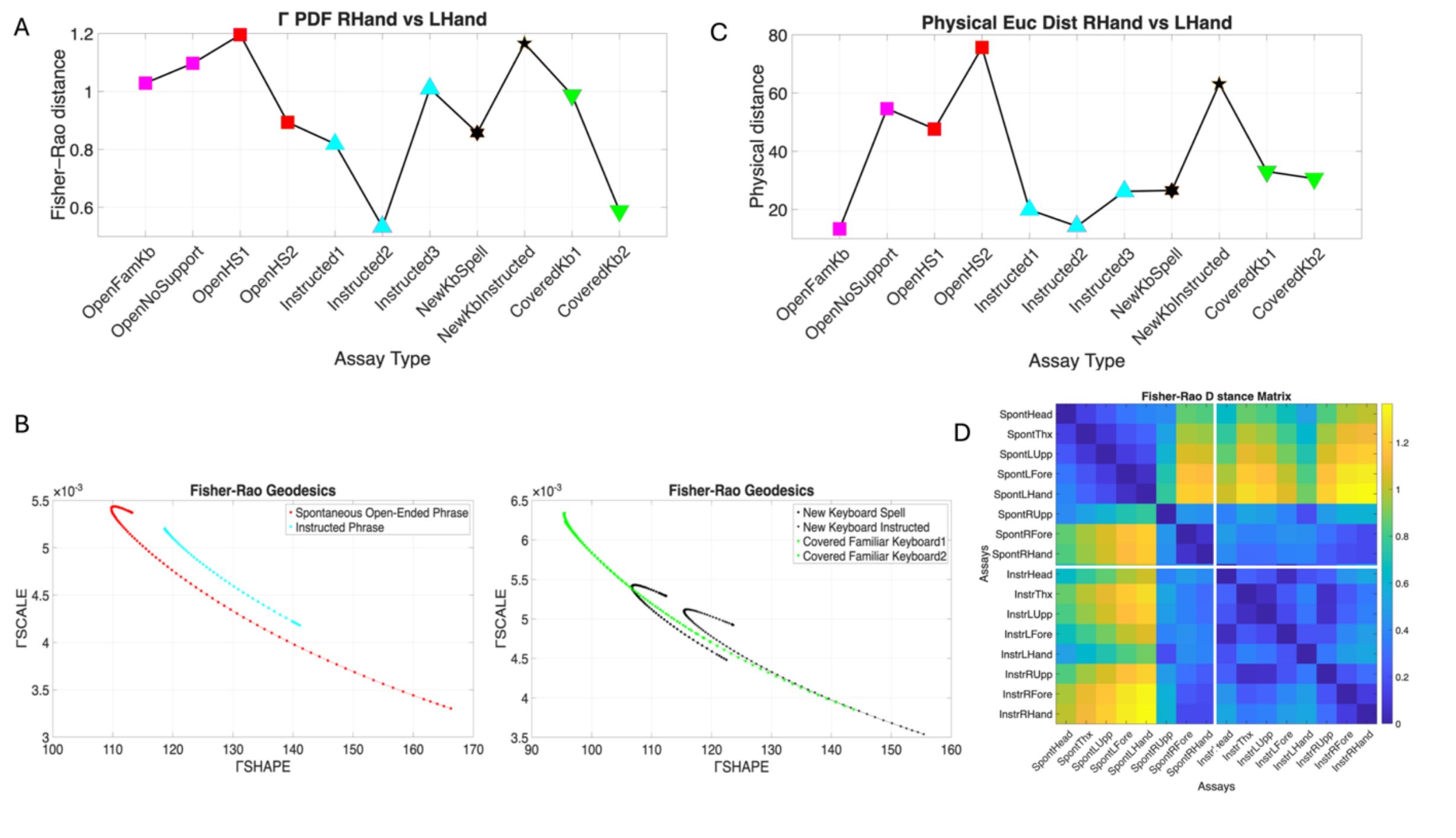

**Figure 13. Geometric analyses of the stochastic Γ manifold trajectories vs. the 3D physical trajectories. (A) Fisher-Rao distance (unitless) between the empirical Γ PDF parameters of the right vs. left hand across all sessions, warmup adaptation and experimental conditions. (B) Geodesics paths of spontaneous open-ended (red) vs. instructed phrase (cyan) left panel and instructed phrase under different cognitive, emotional and memory loads (black new keyboard with new partner in spelling words vs. instructed full phrases; green familiar CRP but covered keyboard). (C) Trends of the physical distance (cm) between the path lengths of the curved movement trajectories described by the right and left hands, as absolute difference between them. (D) Pairwise of the FR-distance across the body, from each body part to each other body part during spontaneous vs instructed conditions. Notice the similarity (close distances between PDFs, blue shades) within the instructed condition vs. the far FR-distances in the dominant hand, forearm of the spontaneous conditions. Notice the far FR-distance as well between all instructed and the spontaneous condition.**

The pairwise FR-distances for the spontaneous and instructed conditions across all 8 sensors demonstrate the differences between these conditions for the dominant vs. non-dominant side of the body. Instructed typing (lower right block) remains more similar (blue hue) across the body while Spontaneous typing to open-ended questions splits with larger differences (yellow hue) across the dominant hand. The cross-condition comparison (upper right and lower left blocks) reveals the large distances during spontaneous typing between the non-dominant side of the body and all sensors across the instructed typing.

### 3.6 Gamma Manifold Fisher-Rao Metric and Geodesic Paths Separate Levels of Cognitive-Memory-Emotional Loads During Instructed Typing

The motoric signatures of instructed typing under variable cognitive loads generated distinct sets of Γ-PDFs. **Figure 8** is high cognitive load (different keyboard, prompter, no haptic support). The communication partner used two forms of prompting to spell, one through phonics and one through spelling the letters. The trajectories capture the continuous flow of the assay. **Figure 9** shows the instructed phrase case, distinguishable in **Figure 12** (black hexagon spelling and black pentagon instructed phrase) for spelling vs. full phrase.

**Figure 13A** shows the FR-distance values of the Γ-PDF differences between the dominant and the non-dominant hands across levels of cognitive loads, while **Figure 13B** shows the corresponding geodesic paths. **Figure 13C** shows the arclength physical distance differences between the hands in correspondence with the FR-distances of panel A.

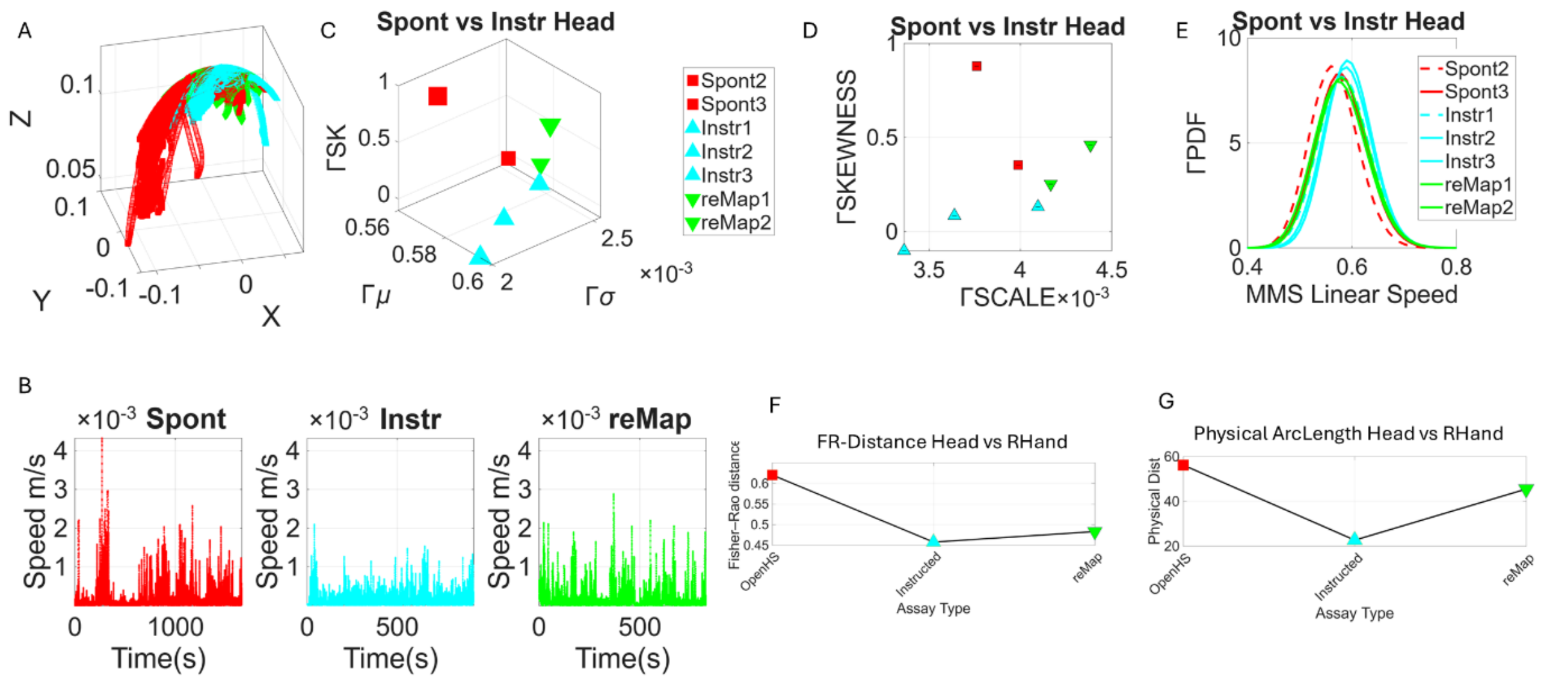

**Figure 14. Head physical and stochastic Gamma trajectories trend together and automatically separate spontaneous responses to open-ended questions from instructed typing the same open-ended phrase, from regression to FC typing under higher cognitive loads. (A) Representative 3D head positional trajectories (meters) for three different conditions: spontaneous typing to answer open-ended question (red); instructed typing of the same phrase responding to open-ended question (cyan); increasing the cognitive load by regressing to facilitated typing with haptic hand support but covered keyboard requiring remapping of visual input to recall letters layout (green). (B) Corresponding linear speed temporal profiles. (C) Γ moments correspond to representative trials of each condition. (D) Corresponding projection of Γ NSR vs. Γ skewness. (E) Empirical Γ PDFs. (F) Fisher-Rao distance from head to the dominant (right) hand typing. (G) Corresponding physical distance as the absolute difference between head and right hand's pathlengths.**

### 3.7 The Head Trajectories are Sufficient to Track the Differences in Both the Gamma Manifold and the 3D Physical Space

The data from the upper body, including the head, thorax, upper arms, forearms and hands revealed unambiguous differences in performance across conditions (Figure 13D). We then asked if the head alone could serve as a proxy to differentiate cognitive effort from authorship. To that end, we examined the FR-distances on the Γ-manifold and on the physical space between conditions and levels of cognitive loads.

**Figure 14** shows the different head motion trajectories captured by the wearable sensor whereby the spontaneous responses to open ended questions occupy a different region of 3D physical space than those from the instructed condition. Furthermore, the excess cognitive load from covering the cardboard and having to remap the keys' layout to type with one finger were also captured in these trajectories. On the Γ-moments space the stochastic signatures grouped differently in spontaneous open ended vs. instructed across different levels of cognitive loads. These differences are expanded in **Supplementary Figure 8** relative to the hands. There we see the head signatures closer to the non-dominant hand. **Supplementary Figure 9** unfolds the FR-geodesics on the plane spanned by the first two Γ-moments. There we see that instructed low vs. high (black markers) span different $\Gamma\mu$ and $\Gamma\sigma^2$ levels along a common geodesic (with higher load red and cyan edges spanning higher variance and lower mean. Lower MMS micro-peaks $\Gamma\mu$ values represent faster raw speed on average (Equation 10). Mild cognitive load in FC with covered keyboard (magenta) sits midway but deviates from the geodesic path. This task bears higher long-term memory load of an old (shared) dyadic motor program. The instructed geodesic paths (black) are different from those spanned by open-ended ones (blue).

**Figure 14D** is a projection plane summarizing the Γ NSR vs. Γ skewness (see also **Supplementary Figure 10**). We appreciate the higher Γ NSR explained by the increase in $\Gamma\sigma^2$ and decrease in micro-peaks $\Gamma\mu$. The decrease in micro-peaks $\Gamma\mu$ corresponds to an increase in average head speed with increase in cognitive load (whereby the denominator in the normalization Equation 10 is larger and as such penalizes the numerator and overall MMS fluctuations). This also impacts the skewness (**Figure 14D**), which shifts positively away from 0 (symmetric case) towards a heavier right tail. This indicates that higher fluctuations of the micro-peaks are not mere outliers but rather more frequent occurrences with increases in cognitive and memory loads. These emerge, respectively, from the keyboard covering, which demands a mental remapping of the keys' layout and from regressing to FC motions of the hand via haptic hand support by the CRP and dyadic exchange of agency levels.

**Figure 14E** shows the shifts in ΓPDFs while **Figures 14F-G** show the congruent changes in FR-distance and Physical head vs. right hand distances. The FR-distance is computed along the geodesics on the Γ manifold whereas the physical distance is taken along the curved arclengths of the head vs. the right-hand movement trajectories in 3D physical space. The trends are similar transitioning from spontaneous phrases in response to open ended questions to deliberate instructed phrases and comparing those to higher cognitive-memory load cases. Each measurement is the average across 11 conditions (4 open ended, 3 instructed, 4 with varying cognitive-memory loads).

### 3.8 Response Latencies Distinguish Cognitive Effort from Semantic Authorship in Open-Ended conditions

Cognitive effort is present in both open-ended questions and instructed-with-cognitive-load conditions. It is important to disentangle the motoric differences as in both cases, the cognitive load leaks into the motor code. What stands out in open-ended questions, where creative thought rather than rote repetition of a prompted phrase drives the system?

Up to this point, we examined the role of the non-dominant side of the body, away from conscious awareness as the dominant side is simultaneously performing the typing action. Here we examine the differences in relation to the performance of the dominant hand. To that end, we look at the latencies of the right-hand motions 1 second from the moment that we asked the open-ended question or gave the instruction of typing the prompted phrase. We did the instructed case with low cognitive load (familiar environment) and mild load (covered keyboard and regression to FC). We highlight in cyan 90 seconds of the performance after the 1 second buffer and compare the motions. All sensors registered data synchronously such that we can simultaneously examine the right dominant hand, left non-dominant hand and the head motion trajectories.

Given the variations across conditions, this approach of allotting a fixed latency for each case made them comparable. It provided enough movements to partially complete the typing task in cases in which starting to type took longer than when typing was immediate. Using techniques from differential geometry of 3D curves, we obtained the Frenet-Serret moving frame. This is a 3D frame of reference composed of the tangent, normal and binormal vectors to the curve. They flow along the curve as the hand translates and rotates, providing metrics of curvature, torsion, speed (arclength) and acceleration of the curve. **Figure 15** provides examples along the trajectories or the dominant, non-dominant hand and the head.

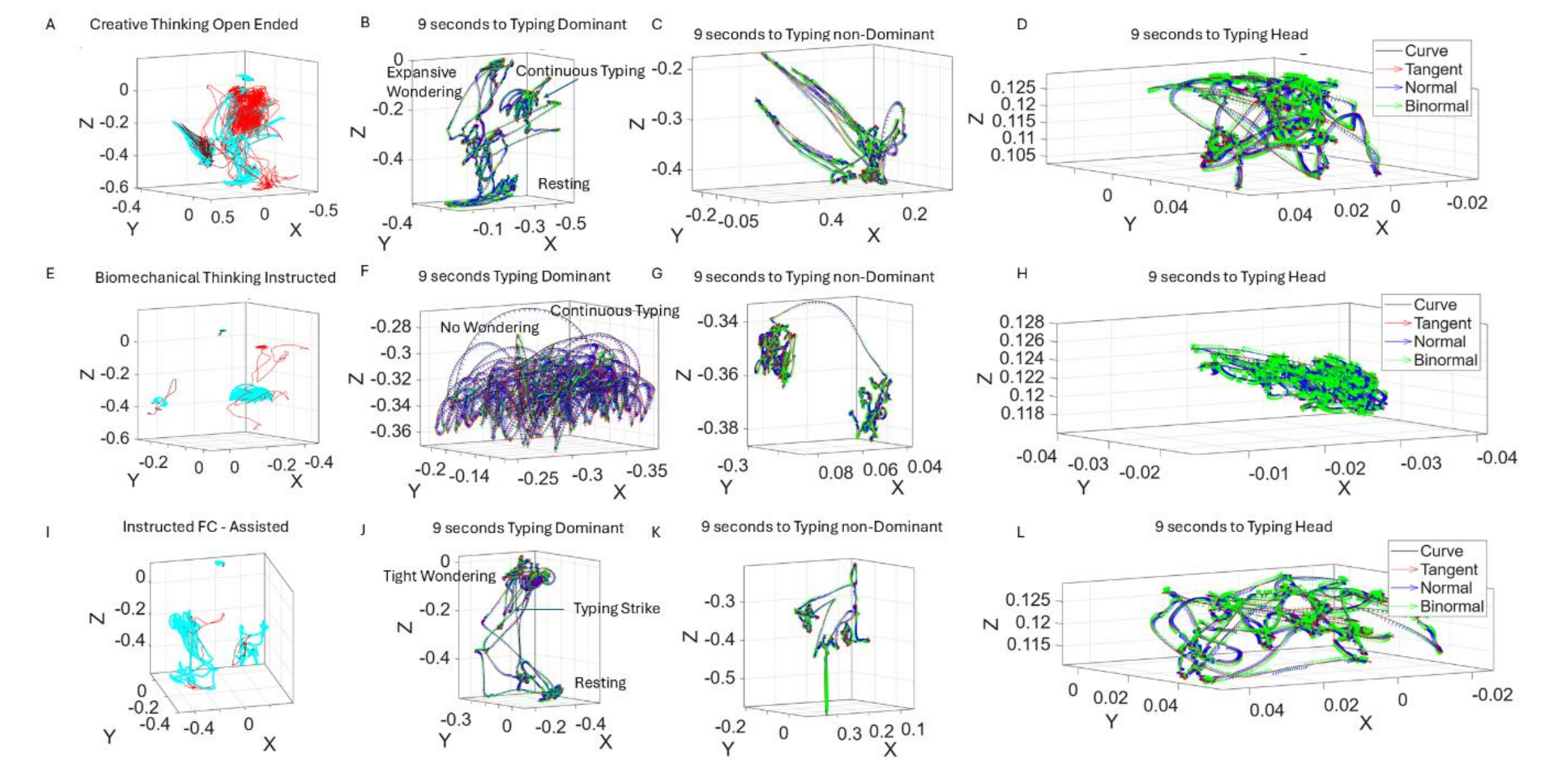

**Figure 15. Latency responses separate semantic authorship from cognitive effort. (A) Spontaneous, creative thinking to answer an open-ended question, with 90 seconds highlighted from one second after asking the question (cyan color). (B) Zooming in the dominant hand movement trajectory for those 9 seconds using the Frenet-Serret moving frame along the trajectory (tangent, normal and binormal vectors giving us curvature, torsion, curve speed and curve acceleration). Notice the structure of the motion parsing space regions into resting the hand below the keyboard, wondering-creative thoughts near the face area, in preparation for an answer and finally the typing on the keyboard. (C) Non-dominant, movements beneath awareness. (D) Head motions. (E) Instructed condition by prompting the same phrase the typer expressed in (A) and marking the paths for 9 seconds from one second after the instruction. This reveals very different typing style. (F) 1 second after the instruction (as in A) the typer typed the phrase without any creative wondering motions. (G) Non-dominant hand shows very different patterns from C and shorter paths (~4cm vs. 40+cm in C). (H) Head motions. (I) FC condition with covered one-finger-strike-board 9 seconds cyan, starting as well one second from instruction and prompting the phrase. (J) Zooming in reveals the typing with one strike per word and doodling thoughts, likely to remap the keyboard and handle the regression to FC. (K) Non-dominant hand motions are shorter than open-ended (~25cm vs. 40+cm in open-ended case). (L) Head motions.**

The latencies of the movements revealed three distinct patterns separating open-ended, low-load instructed and mild-load instructed motions. In open-ended cases, the start of the typing begins much later than in the other two conditions. The participant visibly thinks of the answer while elaborating it in a creative thinking mode. This consists of a preamble of wondering motions of the dominant hand near his face / resting regions and expanding across the peripersonal space, while thinking. We coin these *wondering motions* as they represent cognitive overflow, but their spatio-temporal characteristics isolate the creative from the high-load cognitive states.

In the creative case, the wondering trajectories are expansive. They occupy larger regions of the peripersonal space (**Figure 15B**). These creative, wondering trajectories differ from those of the mild-load condition which remain tightly concentrated in one spot (**Figure 15J**). The wondering motions are altogether absent from the low-load instructed condition using a familiar keyboard, familiar CRP, and familiar prompting style. With lower cognitive demand, the participant goes straight to typing (**Figure 15E-F**) and for 90 seconds he types most of the prompted phrase (**Figure 15F**). In contrast, 90 seconds in the other conditions show his wondering thoughts leaking into the dominant hand movements. The leaking is however distinct and reveals the differences between creating a phrase anew and reorganizing the mental map of the covered keyboard to type that same phrase with FC support. We appreciate the differences in the head trajectories and Frenet-Serret moving frames (15D open-ended, 15H mild, 15L low load). **Supplementary Figure 15** expands on this point and shows for the full latency of 100 seconds, the differences in Γscale vs. units of time and corresponding paths on the $\Gamma\mu$ vs. $\Gamma\sigma^2$ plane.

In the open-ended case, MH types a portion of the phrase upon completing the wondering thoughts (see **Figure15B** arrow pointing to the segment of the typing strikes). That is, he finishes elaborating the creative process and types the new phrase without returning to wondering thoughts. In contrast, during the mild-load instructed case, within the 90 seconds allotted he intermittently strikes the covered board and goes back to wondering motions in tight regions near the face and the resting areas (near the table). He only strikes 3 times the covered board with the FC support and after each strike, he says part of the phrase and returns the hand to the face area where the hand tightly hangs in thought before launching the next FC assisted strike. No expansive wondering occurs prior to typing during mild-low instructed, when requiring remapping of the keyboard and regression to childhood memory of those supported motor programs. Instructed typing with low load has no thinking preamble, just immediate onset of typing motions.

## 4 Discussion

This work proposes a new comprehensive assay and set of biometrics to establish motoric indexes of creative authorship vs. cognitive effort in communication through typing. The results from our case study revealed that the non-dominant part of the body, which moves largely beneath awareness of the typer and of the CRP, funnels out the differences in cognitive load when the typer spontaneously responds to open-ended questions vs. when the same phrase from that spontaneous response is used in deliberate instructed typing by prompting the phrase. These profound differences in the movement trajectories of the non-dominant part of the body consist of a dormant state during instructed biomechanical typing with low cognitive load. They however return to full engagement when we add cognitive load to the deliberate instructed typing. Through the physical movement trajectories in 3D space and through the abstract geodesics of the Γ-manifold, the non-dominant side of the body unambiguously reveals both the type of creative thinking that goes with the open-ended questions and the biomechanical thinking that goes with the typing of the same phrase, when the phrase is prompted. What is then the role of the dominant hand performing the intentional typing motion?

*Semantic Ownership vs. Cognitive Effort:* The role of the dominant hand is evident during creative thinking, prior to and during engaging in the typing of a new phrase. The signatures on the Γ moments space in this condition unfold along different FR-geodesic paths (see also the head signatures in **Figure 14C** and **Supplementary Figure 9**). The spontaneous open-ended vs. deliberate instructed assays can automatically separate conditions using the slopes and intercepts of the scaling power law relations spanned by the Γ-parameters plane and differential entropy. **Figure 15** also illustrates this point zooming into the physical 3D trajectories of the first 90 seconds of the typing motions of the same phrase in **Figure 7**. Detailed geometric analyses of the 3D curves involving the Frenet-Serret moving frames quantified such semantic ownership in both the dominant (intended) and non-dominant (spontaneous motions) as they unfold simultaneously in each assay. Whereas the open-ended creative thoughts unfold in the first 90 seconds of the motion through wondering of the dominant hand, the instructed movements nearly completed typing the phrase with no wondering prelude.

*Ownership in FC*: Regressing to childhood FC-style, we examined a standardized index of motor agency as a ratio of ΓSNR for each assay on the denominator, and the sum of ΓSNR plus the level of Approximate Entropy of each assay in the denominator. As Approximate Entropy increases and the motions are unpredictable, Agency decreases. We found that during FC the CRP levels of Approximate Entropy overpowered her levels of agency. As her agency level decreased, she must have surrendered her control in favor of passive support to the typer's hand. In contrast, the typer's sensor on the forearm leading the hand during this form of assisted typing had higher levels of agency and more predictive value of Approximate entropy (lower values). This was also attested by the relationship between our normalized index of Agency and the Gamma differential entropy. Together with the (non-obvious) patterns of the non-dominant side of the body broadcasting the levels of spontaneity or deliberateness, or levels of cognitive loads during typing, these biometrics provide new ways to track the progression of the learning process that takes place in communication through typing. This is important as harsh criticism of supported communication has stalled or altogether halted the teachings of these methods requiring haptic support in the initial stages of learning to move and control the eye-head-trunk-arm-hand coordination that must take place for the pointing to succeed and acquire the target in a timely fashion.

## 4.1 A New Path for Supported Typing

Biometrics and assays here offer a new avenue for learning to type with support because they are agnostic and objective. They unambiguously reveal the level of agency of the typer and the CRP. The participant in this study is an advanced typer who can vocalize what he types, and, in this way, his performance offers a way to verify the message directly. However, the assays and biometrics presented here are general and work with any level of communication and any type of typing style. A novice who still needs haptic support from a communication partner, whether holding a hand or holding a board where the typer spells to communicate, or pointing to a piece of paper, etc. can also provide motoric data produced under the assays in **Figure 3**. Combining the measures of the non-dominant side of the body and those of the typing arm and the supporting hand give a full package to track levels of agency, autonomy and control in the dyad. Should the CRP provide excessive support and decrease the typer's agency index, we can alert the CRP to back off and allow the typer to lead. Metrics of transfer entropy, approximate entropy, Γ differential entropy can give us this ability. As such it is OK to follow the process of learning to type with appropriate support.

We caution that our biometrics and assays do not say that any given person owns the typed message. They merely point out the degree to which the motions are tied to cognitive effort or creative thoughts. A novice will first have to learn to own their actions, to self-regulate them, and to fluidly navigate the intrinsic and extrinsic sensory-motor sources of their own body in motion. These biometrics and assays will help the novice get there because they measure performance objectively and agnostically, in a personalized manner. They measure the person's performance to protect the person from excessive control by others and to nurture motoric communication towards full autonomy. By revealing who leads with agency in this dyadic interaction we offer the possibility of accelerating the learning process in a respectful and humane manner.

The case of MH gave us an upper bound to measure departure from and help others start this journey with concrete higher-level goals. We know the destination and can now have rigorous methodology to move in that direction. This work is an invitation to join us and work together with an unprecedented combination of rigor, transparency and kindness. As altricial mammals with a long developmental path at birth, we all started with assistance from our progenitors. It is nature's way to help us grow strong but it is also, most importantly, our right because we are alive.

## 4.2 A Useful Analogy Moving Forward

The process of learning to type with support is akin to the process of Brain Computer Interfaces (BCIs) or Brain Machine Interfaces (BMIs) where the brain signals are used to learn to control an external device (e.g., a cursor on a computer screen [53] or a robotic arm [72, 73]). In those learning-with-assistance scenarios a computer algorithm plays the role of the CRP, and the brain signals represent the pointing hand of the typer. In the initial stages of assisted typing, the CRP heavily guides the hand of the typer to the desired letters. The typer's contribution is minimal to null. This is much like the algorithm, rather than the brain signals, moving the cursor or the robotic arm, in the initial stages of learning, to calibrate the feedback control loop. In due time, the CRP relinquishes a bit of this control over the typer's hand and lets the typer learn from the errors. Likewise, the BMI/BCI algorithm lets the brain signals do some of the control of the cursor or the robotic arm, and the brain learns from its own errors. Eventually, one has 50-50 control of the algorithm and brain signal (a flip of a coin) controlling the cursor or the robotic arm is random; then 60-40 then 70-30, etc. Eventually, the brain signals gain 100% control of the external device and learn to control it with full agency. Likewise, CRPs can let go 100% of their control and hand it over to the typer who acquires full agency, even when the letter board may still be held by the CRP or the hand may still need support. Passive support where the CRP has low agency and the typer controls the flow is not only possible (as we saw in this example), but also most importantly, it is quantifiable in multiple reliable ways that do not depend on the biased opinion or agenda of an observer. We have a new path forward for typers who inevitably need haptic support.

### 4.3 Stochastic Equivalence of Physical Performance

The stochastic changes derived from the bodily movements were appropriately measured by the FR-distance metric on the curved Γ parameters manifold and, in the 3D, physical space. There is congruence between these changes according to the distance metrics on the abstract Γ manifold and the physical spaces. In both spaces there is a tendency to increase the distance between the dominant and non-dominant hands with cognitive load and to decrease it with instruction. Yet, under instruction and higher cognitive-memory-emotional loads, the distances tend to increase again. These changes, which we also anchored to the non-dominant hand demonstrate a new and non-obvious way to quantify cognitive load using the motoric signatures in ways that cannot be intentionally controlled by an outsider. In this sense, they offer a new way to measure authorship of our cognitive thoughts, during spontaneous elaboration and under prompted instruction.

### 4.4 Utility and Scalability: Head Motions as Proxy of the Body

The utility of this finding is linked to the head movements. Head motions alone capture these differences in cognitive load and intent. As such, they provide a proxy of their motoric signatures across the body capturable with simple unobtrusive means, such as video. Because of the disparity in levels of mental awareness quantified between the dominant and non-dominant sides of the body, and because of the fidelity of trends in physical changes and trends in Γ stochastic changes, we propose that both the stochastic and physical signatures of cognitive load presented in this paper as indexes of degree of personalized authorship can be transferred and scaled to other cases.

The suite of assays that we offer can be adapted to individuals who do not talk, such that spontaneous gestures and any spontaneous vocalization capturable through video and mics can be used to probe the system in deliberate instructed mode. These finer motor vocalization signals are nevertheless motorically quantifiable. They may give us a window into other issues with smooth and striated muscles across systems. By contrasting spontaneity and deliberateness in our gestures, voice and facial expressions, we can separate motoric signatures of cognitive loads from motoric signatures of mere biomechanical performance. This opens a new way to quantify in the non-dominant end-effector of the person, the individualized index of communication's authorship. No one other than the typer's brain itself could reproduce such geodesics on the Γ manifold in correspondence with those of his movements in 3D physical space. The dominant hand then adds a unique, personalized, unambiguous signature of ownership of the person's spontaneous creative thoughts preluding the typing action in ways unique to the person. They are a large departure from what precedes the typing of the instructed, prompted phrases -even of those instructed with varying degrees of cognitive and memory effort.

### 4.5 Space-Time Equivalence Through Geodesics

One of the most compelling results of our analyses was the equivalence of temporal and spatial features which manifested both in the geodesics of the Γ manifold and in the arclength and time delays of the hand's trajectories. As we obtained the motion trajectories across conditions within a fixed time length (**Figure 15 A-E-I**) we extract the amount of space traveled and the latencies of each case. They match those of the Γ FR-geodesics across conditions (**Figure 13, 15** and **Supplementary Figure 15**). Their differences provide a metric encompassing both the variations in physical space corresponding to the variations in Γ space. The latter offering the ability to capture more general variations of micro-peaks in other biorhythms co-registered from the nervous systems (**Supplementary Figure 4**). In this sense, this outcome is what enables us to generalize these results to other end effectors and signals, such that we can use other end effectors (tongue, eyes, face, head) to acquire the spatial and temporal signatures distinguishing across cognitive effort, biomechanical effort and creative thinking preambles to the onset of typing.

### 4.6 Caveats, Path to Generalization and Conclusions

This is a case study of an advanced typer with emerging spoken language. It serves as our upper bound to track and verify performance in general. We need to scale the study and sample broadly across different skill levels of communicating through typing or pointing in the absence of spoken language. Our methods offer a way to profile the somatic-sensory-motor systems to determine where to start based on regulation level requirements (**Supplementary Figure 4**). Today we can do this unobtrusively and at scale with our Lab in the Box approach [14].

Although there are idiosyncratic aspects of MH performance that pertain to his personal style of typing, self-regulating, learning trajectory, and acquired skills, the assays and biometrics that we present are universal. They can be adapted to assess learning in a novice typer, who cannot talk and is learning to type with assistance. They can also evaluate more advanced, independent typers who may not need as much support but cannot talk either. They can be adapted to individuals who communicate through echolalia or self-talks/vocalization, using the same basic assay and biometrics. We acknowledge that this work nonetheless must be carried out to both reveal broad ranges across the population of non-speakers and to help accelerate their learning process. But we also provide the path forward, moving away from behaviorism into a humane approach that provides agency, while helping the person self-regulate and be comfortable.

Sensory-motor variability and behavioral performance during communication are as heterogeneous as the ns-ASD population and as any other non-speaking group (stroke, CP, etc.) These differences are nonetheless quantifiable with the methods introduced here. These methods remain agnostic to the messaging or authorship claims. They can objectively reveal through precise spatiotemporal biometrics who leads with agency in a dyadic setting and because of the stochastic and geometric rules simplifying computational complexity through important parameter identification, they can be embedded in cell phones and broadly scaled. We underline the need to extend this study to other groups and offer the metrics to safely differentiate levels of cognitive effort during instructed typing from motions that are preamble to typing creative thoughts.

## 5 Conflict of Interest

The author declares that Rutgers University and the author share patents associated with this work. These patents could result in future financial benefits to the university and the author.

## 6 Author Contributions

Conceptualization EBT, Data curation EBT, Formal analysis EBT, Funding acquisition EBT, Investigation EBT, Methodology EBT, Project Administration EBT, Software, Visualization and Validation EBT, Writing original draft EBT.

## 7 Funding

This work was funded by the Nancy Lurie Marks Family Foundation Career Continuation Award to EBT and the Rutgers University Social Arts and Science External Funds for Community Engagement Award to EBT.

## 8 Acknowledgments

We thank the Nancy Lurie Marks Family Foundation for supporting our work on Autism. We thank Matthew Hayes and Rene Hayes for their collaboration. We thank all members and visitors to the Sensory Motor Integration Laboratory for their continuous help as we develop this research program.

## 9 Tables

Table 1. Assay to track authorship levels.

| **Cognitive load** | **Low** | **Low** | **Mid** | **Mid** | **High** |
|---|---|---|---|---|---|
| **Assay Used** | Open Ended Q,<br>Familiar keyboard,<br>Familiar CRP,<br>Familiar prompter, | Instructed Phrase,<br>Familiar KB,<br>Familiar CRP,<br>Familiar prompter, | Open-ended Q,<br>Familiar KB,<br>Familiar CRP,<br>Familiar prompter,<br>Regression to FC | Open-ended Q,<br>Familiar KB,<br>Covered KB, Familiar CRP,<br>Familiar Prompter,<br>Regression to FC | Open-ended Q, Spelling words by phonics vs letters,<br>New KB,<br>New CP,<br>Unfamiliar prompter |
| **Saved Data Streams** | Video/Biosensors | Video/Biosensors | Video/Biosensors | Video/Biosensors | Video/Biosensors |

# 11 Data Availability Statement

The datasets GENERATED/ANALYZED for this study can be found in the ZENODO [https://zenodo.org/records/24432280] **Doi 10.5281/zenodo.24432280** .